\documentclass[12pt]{article}
\usepackage{amsmath}
\usepackage{amssymb}
\usepackage{multirow}

\makeatletter\@addtoreset{equation}{section}
\makeatother
\allowdisplaybreaks[1]
\begin{document}
\begin{titlepage}
\begin{flushright}
TIT/HEP-645 \\
CTP-SCU/2015019 \\
September 2015
\end{flushright}
\vspace{0.5cm}
\begin{center}
{\Large \bf
BPS Equations in $\Omega$-deformed $\mathcal{N}=4$ Super Yang-Mills Theory
}
\lineskip .75em
\vskip0.5cm
{\large Katsushi Ito${}^{1}$, Yusuke Kanayama${}^{1}$, Hiroaki Nakajima${}^{2}$ and Shin
Sasaki${}^{3}$ }
\vskip 2.5em
${}^{1}$ {\normalsize\it Department of Physics,\\
Tokyo Institute of Technology\\
Tokyo, 152-8551, Japan} \vskip 1.0em
${}^{2}$ {\normalsize\it Center for Theoretical Physics,\\
Sichuan University\\
Chengdu, 610064, China} \vskip 1.0em
${}^{3}$ {\normalsize\it Department of Physics,\\
Kitasato University\\
Sagamihara, 252-0373, Japan}
\vskip 3.0em
\end{center}
\begin{abstract}
We study supersymmetry 
 of ${\cal N}=4$ super Yang-Mills theory in four dimensions deformed in the
$\Omega$-background. We take the Nekrasov-Shatashvili limit of the background 
so that two-dimensional super Poincar\'e symmetry is recovered.
We compute the deformed central charge of the superalgebra and study the 1/2 and 1/4 BPS states. 
We obtain the $\Omega$-deformed 1/2 and
 1/4 BPS dyon equations from the deformed supersymmetry
 transformation and the Bogomol'nyi completion of the energy.
\end{abstract}
\end{titlepage}

\baselineskip=0.7cm
\tableofcontents%
\section{Introduction}

The $\Omega$-deformation of supersymmetric gauge theories has been
studied extensively since it gives a very useful regularization to calculate the 
non-perturbative instanton corrections to the partition functions \cite{Moore:1997dj,Nekrasov:2002qd}.
This deformation has been developed for theories with ${\cal N}=2$ supersymmetry in four dimensions \cite{Losev:2003py,Nekrasov:2003rj,Ito:2010vx}. 
It has been generalized to the ${\cal N}=4$ super Yang-Mills
theory in four dimensions \cite{Ito:2011cr,Ito:2012hs,Hellerman:2012rd} and also 
theories in various dimensions \cite{Shadchin:2006yz, Ne-M, Dimofte:2010tz, Reffert:2011dp}.

The $\Omega$-background is realized as a spacetime fibration over
a torus with the actions of an abelian group.
This is also obtained by the dimensional reduction from higher
dimensional curved spacetime compactified on torus. 
For this reduction it is necessary to introduce  the  R-symmetry Wilson line gauge fields along the torus in order to 
recover a part of supersymmetry.
The R-symmetry Wilson lines are identified with the contorsion along the
torus \cite{Ito:2012hs}.
This deformation however breaks Poincar\'e symmetry in general and the deformed supersymmetry algebra
takes the form of nilpotent type, which means that  the BPS spectrum is not well-defined. 

 The Nekrasov-Shatashvili (NS) limit of the $\Omega$-background is the limit such that a part of the torus action is removed \cite{Nekrasov:2009rc}. 
 In this limit we recover the translational symmetry in the subspace and the
 supersymmetry algebra is enhanced to the super Poincar\'e algebra.
 Then  one can study the BPS spectrum based on the deformed supersymmetry.
Moreover in the NS limit the effective two-dimensional theory has the non-trivial vacuum structure related to the integrable models \cite{Nekrasov:2009rc,Nekrasov:2010ka,Bulycheva:2012ct}. 
It will be an interesting problem how such integrable structure appears from the UV point of view. 
The  BPS solitons would play an important role to study this integrable structure.
 
 For ${\cal N}=2$ supersymmetric Yang-Mills theory, the deformed BPS equations were studied in
 \cite{Ito:2011ta}. 
In particular the deformed monopole equations can be solved by the B\"acklund
 transformations   as the undeformed one \cite{Forgacs:1980yv}. 
 Moreover in \cite{Bulycheva:2012ct} (see also \cite{Tong:2015kpa}), they proposed the non-trivial vortex solution induced by the $\Omega$-background.
The purpose of this paper is to
 study deformed BPS states for $\Omega$-deformed ${\cal N}=4$ super Yang-Mills theory. 
This is also interesting because these BPS states play an important role for understanding S-duality of the $\Omega$-deformed gauge theories.

In this paper we focus on the 1/2 and 1/4 BPS dyon equations in ${\cal N}=4$ supersymmetric Yang-Mills theory in the NS
limit of the $\Omega$-background.
In a previous paper \cite{Ito:2013eva} three of the present authors  studied the deformed supersymmetry
algebra in the $\Omega$-background as well as its NS limit.
There are three types of $\Omega$-deformations which are associated with three types
of topological twist of ${\cal N}=4$ supersymmetry \cite{Yamron}.
These are the half (or Donaldson-Witten \cite{Witten:1988ze}), the Vafa-Witten \cite{Vafa:1994tf} and  the Marcus (or generalized Langlands or Kapustin-Witten) \cite{Marcus:1995mq,Kapustin:2006pk} twists.
 In the NS limit, the $\Omega$-deformations 
for the 
Vafa-Witten and the Marcus 
twists
are shown to be equivalent by 
 some identification. 
 It  turns out that only the two types of deformations are independent. 
One of  the new and interesting properties in ${\cal N}=4$ theory
is the $1/4$ BPS states 
\cite{Bergman:1997yw,Hashimoto:1998zs,Kawano:1998bp,Bergman:1998gs,Hashimoto:1998nj,Lee:1998nv,Houghton:1999cm}.
In this paper we will find the 
deformed $1/4$ BPS equations 
for the Vafa-Witten twist.
The BPS equations are similar to the vortex type equations and might have vortex type central
charge  \cite{Bulycheva:2012ct}. 
We will calculate the central charge carefully from the deformed supersymmetry algebra and
show that there is no vortex type central charge in ${\cal N}=4$ theory.

This paper is organized as follows:
in section 2, we review the $\Omega$-deformation of $\mathcal{N}=4$ super Yang-Mills theory.
In section 3, we study deformed supersymmetry algebra in the NS limit and calculate the central charges.
In section 4, we argue the BPS equations from the superalgebra and also the energy bound.
Section 5 is devoted to conclusions and discussion.
In appendix A, we summarize the notation of the Dirac matrices in four and six dimensions.
In appendix B, we investigate the vacuum conditions based on the $\Omega$-deformed Lagrangian.

\section{
$\mathcal{N} = 4$ super Yang-Mills theory in $\Omega$-background
}
\subsection{Setup}

In this section, 
we introduce four-dimensional $\mathcal{N} = 4$ super Yang-Mills theory with gauge group $G$ in 
$\Omega$-background.
The theory is obtained by the dimensional reduction of the ten-dimensional
$\mathcal{N} = 1$ super Yang-Mills theory in the $\Omega$-background
metric with torsion \cite{Ito:2012hs}.
We start from the $\mathcal{N} = 1$ super Yang-Mills theory 
in general curved background with Euclidean signature\footnote{In this section we use
Euclidean signature to define the $\Omega$-background. In the next section,
we perform the Wick rotation and use the Minkowski signature to
study the BPS equations.}. We denote $x^{\mathcal{M}}$ ($\mathcal{M}=1,\ldots,10$) as spacetime coordinates.
The metric $g_{\mathcal{M}\mathcal{N}}$ is written in terms of the vielbein $e^M{}_\mathcal{M}$ as
$g_{\mathcal{M}\mathcal{N}}=\eta_{MN}e^M{}_{\mathcal{M}}e^N{}{}_{\mathcal{N}}$,
where 
$\eta_{MN}=\mathrm{diag}\,
(+1,+1,\ldots,+1)
$.
The capital letters $M, N, P, \ldots = 1, \ldots, 10$ are indices of the
tangent space coordinate.
The ten-dimensional
$\mathcal{N} =1$ vector multiplet consists of the gauge field
$A_{\mathcal{M}}$ and the $SO(10)$ spinor field $\Psi$, which belong to
the adjoint representation of the gauge group $G$.
The action is
\begin{align}
S_{10D}
&=
\frac{1}{\kappa g_{10}^{2}}
\int \! d^{10} x \ 
\mathrm{Tr}\biggl[
\frac{1}{4}e\bigl(e^{\mathcal{M}}{}_{M}e^{\mathcal{N}}{}_{N}
\widehat{F}_{\mathcal{M}\mathcal{N}}\bigr)^{2}
+\frac{i}{2}e\,\bar{\Psi}\Gamma^{\mathcal{M}}
\widehat{\nabla}_{\mathcal{M}}^{(G)}\Psi
\biggr].
\label{curvedlag}
\end{align}
Here $\kappa$ is the normalization factor for the generators of the gauge group 
 and $g_{10}$ is the gauge coupling constant.
$e^{\mathcal{M}}{}_M$ is 
the inverse of $e^M{}_{\mathcal{M}}$ and $e = \det e^M{}_{\mathcal{M}}$.
The field strength $\widehat{F}_{\mathcal{M}\mathcal{N}}$ of the gauge field $A_{\mathcal{M}}$ in the background 
with the torsion $T_{\mathcal{MN}} {}^{\mathcal{P}}$ is defined by 
$
 \widehat{F}_{\mathcal{M}\mathcal{N}}=F_{\mathcal{M}\mathcal{N}}
-T_{\mathcal{M}\mathcal{N}}{}^{\mathcal{P}} A_{\mathcal{P}}
$, 
where $F_{\mathcal{M}\mathcal{N}}$ is the ordinary gauge field strength
$F_{\mathcal{M}\mathcal{N}}=
\partial_{\mathcal{M}}A_{\mathcal{N}}-\partial_{\mathcal{N}}A_{\mathcal{M}}
+i[A_{\mathcal{M}},A_{\mathcal{N}}]$.
The Dirac matrices $\Gamma^M$ are defined by
the relation $\Gamma^M\Gamma^N+\Gamma^N\Gamma^M=2\eta^{MN}$ and
$\Gamma^{\mathcal{M}}=e^{\mathcal{M}}{}_{M}\Gamma^{M}$.
The gauge covariant derivative is defined by 
$\widehat{\nabla}^{(G)}_{\mathcal{M}}\ast=\widehat{\nabla}_{\mathcal{M}}\ast+i[A_{\mathcal{M}},\ast]$,
where 
$\widehat{\nabla}_{\mathcal{M}}$ acts on the spinors as 
$
 \widehat{\nabla}_{\mathcal{M}}\Psi=\left(\partial_{\mathcal{M}}+\frac{1}{2}
\widehat{\omega}_{\mathcal{M},NP}\Gamma^{NP}\right)\Psi
$.
Here 
$\Gamma^{MN}=
\frac{1}{4}(\Gamma^M\Gamma^N-\Gamma^N\Gamma^M)
$
is the $SO(10)$ local Lorentz generator and 
$\widehat{\omega}_{\mathcal{M}, NP}$ is the spin connection including the torsion.

We will consider the invariance of the action under
 the following supersymmetry transformation:
\begin{align}
\delta_{\zeta} 
A_{\mathcal{M}}
=
i\bar{\zeta}\,\Gamma_{\mathcal{M}}\Psi, \quad
\delta_{\zeta}
\Psi
=
- \widehat{F}_{\mathcal{M} \mathcal{N}} \Gamma^{\mathcal{M} \mathcal{N}} \zeta,
\label{SUSYtransf}
\end{align}
where the supersymmetry parameter
$\zeta$ satisfies the parallel spinor condition 
\begin{align}
\widehat{\nabla}_{\mathcal{M}} \zeta = 0.
\label{parallel10D}
\end{align}
If the action is invariant under the transformation \eqref{SUSYtransf},
we can derive the supersymmetry algebra 
from the supersymmetry transformation of the
supercurrent \cite{Osborn:1979tq} (see also \cite{Yokoyama:2015yga}).
From the Noether procedure
we have the supercurrent with the parameter $\zeta$ as 
\begin{align}
j^{\mathcal{M}}_{\zeta} =
\frac{1}{\kappa g_{10}^2}\mathrm{Tr}\,
\left[
ie\widehat{F}_{\mathcal{N}\mathcal{P}}\bar{\Psi}
\Gamma^{\mathcal{M}}\Gamma^{\mathcal{NP}}\zeta\right].
\label{eq:supercurrent}
\end{align}
Its supersymmetry transformation with the parameter $\xi$ is calculated as
\begin{align}
\delta_{\xi}j^{\mathcal{M}}_{\zeta}
&=
\frac{1}{\kappa g_{10}^{2}}\mathrm{Tr}\biggl[
2i\bar{\xi}\Gamma^{\mathcal{N}}\zeta e\biggl(
\widehat{F}^{\mathcal{MP}}\widehat{F}_{\mathcal{NP}}
-\frac{1}{4}\delta^{\mathcal{M}}{}_{\mathcal{N}}
\widehat{F}^{\mathcal{PQ}}\widehat{F}_{\mathcal{PQ}}
+\frac{1}{2}\bar{\Psi}\Gamma^{\mathcal{M}}\widehat{\nabla}^{(G)}_{\mathcal{N}}
\Psi
+\frac{1}{2}\bar{\Psi}\Gamma_{\mathcal{N}}\widehat{\nabla}_{(G)}^{\mathcal{M}}
\Psi
\biggr)
  \notag\\[2mm]
 &\qquad\qquad{}
-\frac{e}{2}\bar{\xi}\Gamma_{\mathcal{P}}\zeta\ 
\widehat{\nabla}^{(G)}_{\mathcal{N}}
\bigl(\bar{\Psi}\Gamma^{\mathcal{MNP}}\Psi\bigr)
+\frac{e}{4}\bar{\xi}\Gamma^{\mathcal{NPQ}}\zeta\ 
\widehat{\nabla}^{(G)}_{\mathcal{R}}
\bigl(2\delta^{\mathcal{R}}_{\mathcal{Q}}
\bar{\Psi}\Gamma^{\mathcal{M}}{}_{\mathcal{NP}}\Psi
-\delta^{\mathcal{M}}_{\mathcal{Q}}
\bar{\Psi}\Gamma^{\mathcal{R}}{}_{\mathcal{NP}}\Psi
\bigr)
 \notag\\[2mm]
 &\qquad\qquad{}
+\frac{i}{4}\bar{\xi}\Gamma^{\mathcal{MNPQR}}\zeta e
\widehat{F}_{\mathcal{NP}}\widehat{F}_{\mathcal{QR}}
 \biggr],
\label{currenttrdef}
\end{align}
where we have used the equation of motion for $\Psi$. 
$\Gamma^{\mathcal{MNP}}$ and 
$\Gamma^{\mathcal{MNPQR}}$ are the totally antisymmetrized products of 
the Dirac matrices with the weights $\frac{1}{3!}$ and $\frac{1}{5!}$,
respectively.
The supercharge $\mathcal{Q}$ is defined by 
$\bar{\zeta}\mathcal{Q}=\int d^{9}x\,j^{\sharp}_{\zeta}$,
where $j^{\sharp}_{\zeta}$ is the temporal component of the supercurrent, and $d^9x$ is the spatial volume element.
Since the left hand side of \eqref{currenttrdef} can be rewritten as 
$[i\bar{\xi}\mathcal{Q},\,j^{\mathcal{M}}_{\zeta}]$,
the supersymmetry algebra is obtained by the spatial integration of the 
temporal component of \eqref{currenttrdef} as 
\begin{align}
[\bar{\xi}\mathcal{Q}, \bar{\zeta}\mathcal{Q}]
=2\bar{\xi}\Gamma^{M}\zeta\ P_{M}
+\frac{1}{4}\bar{\xi}\Gamma^{MNPQR}\zeta\ 
\mathcal{Z}_{MNPQR}\ ,
\label{SUSYalg10D}
\end{align}
where $P_{M}=\int d^{9}x\,e^{\mathcal{N}}{}_{M}
\mathcal{T}^{\sharp}{}_{\mathcal{N}}$ is 
the ten-dimensional momentum.
$\mathcal{T}^{\sharp} {}_{\mathcal{N}}$ is the temporal component of the
energy momentum tensor $\mathcal{T}^{\mathcal{M}}{}_{\mathcal{N}}$ 
, which is given by 
\begin{gather}
\mathcal{T}^{\mathcal{M}}{}_{\mathcal{N}}
=
\frac{1}{\kappa g_{10}^{2}}\mathrm{Tr}\biggl[
e\widehat{F}^{\mathcal{MP}}\widehat{F}_{\mathcal{NP}}
-\frac{e}{4}\delta^{\mathcal{M}}{}_{\mathcal{N}}
\widehat{F}^{\mathcal{PQ}}\widehat{F}_{\mathcal{PQ}}
+\frac{e}{2}\bar{\Psi}\Gamma^{\mathcal{M}}
\widehat{\nabla}^{(G)}_{\mathcal{N}}\Psi
+\frac{e}{2}\bar{\Psi}\Gamma_{\mathcal{N}}
\widehat{\nabla}_{(G)}^{\mathcal{M}}\Psi
\biggr]. 
\label{EMtensor10D}
\end{gather}
The central charge $\mathcal{Z}_{MNPQR}$ is defined by 
\begin{gather}
\mathcal{Z}_{MNPQR}=\int d^{9}x\, 
\frac{1}{\kappa g_{10}^{2}}\mathrm{Tr}\left[\frac{1}{5!}
\varepsilon_{MNPQR}{}^{STUVW}
ee^{\sharp}{}_{S}\widehat{F}_{TU}\widehat{F}_{VW}
\right],
\end{gather}
where 
$\varepsilon_{M_{1}\cdots M_{10}}$ is the 
totally antisymmetric tensor in ten dimensions and 
$\widehat{F}_{MN}=e^{\mathcal{M}}{}_{M}e^{\mathcal{N}}{}_{N}
\widehat{F}_{\mathcal{MN}}$.
Now we introduce the $\Omega$-background.
The $\Omega$-background metric in ten dimensions is defined by the ${\bf R}^4$ fibration over the torus $T^6$:
\begin{align}
&ds^2=(dx^m+\Omega^m {}_{a} dx^{a+4})^2 + (dx^{a+4})^2,
\notag  \\
&\Omega^m {}_a = \Omega^{mn} {}_{a} x_n, 
\quad \Omega^{mn} {}_{a} = - \Omega^{nm} {}_{a},
\label{Omegabcgdmetric}
\end{align}
where $x^m$ ($m=1,2,3,4$) and $x^{a+4}$ ($a=1,\ldots,6$) are the
spacetime and the internal space coordinates.
The antisymmetric matrices $\Omega_{mna}$ 
are parameterized as 
\begin{align}
 \Omega_{mna}=\left(
\begin{array}{cccc}
0 & \epsilon^1_{a} & 0 & 0 \\
-\epsilon^1_{a} & 0 & 0 & 0 \\
0 & 0 & 0 & -\epsilon^2_{a} \\
0 & 0 & \epsilon^2_{a} & 0 \\
\end{array}
\right),
\label{eq:Omega_matrix}
\end{align}
where $\epsilon^1_{a}$, $\epsilon^2_{a}$ are real constant parameters.
The matrices satisfy the following commutation relation,
\begin{align}
 \Omega_m {}^p {}_a \Omega_{pnb}- \Omega_m {}^p {}_b \Omega_{pna}=0.
\end{align}
We now perform the dimensional reduction of the action \eqref{curvedlag}
with the $\Omega$-background metric \eqref{Omegabcgdmetric} and the
constant torsion, which has nonzero components only in the internal part.
We decompose the ten-dimensional gauge field as
$A_{\mathcal{M}}=(A_{m},\varphi_{a}^{})$, 
where $A_{m}$ is the gauge field and 
$\varphi_{a}$ are the scalar fields in four dimensions. 
The spinor field $\Psi$ is decomposed as 
$\Psi=(\Lambda_{\alpha}{}^{A},\bar{\Lambda}^{\dot{\alpha}}{}_{A})$,
where $\alpha, \dot{\alpha} = 1,2$ and $A = 1, \ldots,4$ are the
four-dimensional $SO(4)$ spinor and $SO(6)_I \simeq SU(4)_I$ R-symmetry
indices, respectively.
Then the action of the four-dimensional $\mathcal{N} = 4$
super Yang-Mills theory in the $\Omega$-background is
\cite{Ito:2012hs}
\begin{align}
S = \frac{1}{\kappa g^2} 
\int \! d^4 x \ 
\mathrm{Tr} 
\Big[ 
 &\, \frac{1}{4} F^{mn} F_{mn} 
+ \frac{1}{2} G^{ma} G_{ma}
+ \frac{1}{4} H^{ab} H_{ab}
\notag \\
 &\, 
+ \Lambda^A \sigma^{m} D_{m} \bar{\Lambda}_A 
- \frac{1}{2} (\Sigma_a )^{AB} \bar{\Lambda}_A [ \varphi_a ,
 \bar{\Lambda}_B ] 
- \frac{1}{2} (\bar{\Sigma}_a )_{AB} \Lambda^A [ \varphi_a , \Lambda^B ] \notag \\
 &\, - \frac{i}{2} \Omega^{m}_a \big( ( \Sigma_a )^{AB}
 \bar{\Lambda}_A 
D_{m} \bar{\Lambda}_B + (\bar{\Sigma}_a )_{AB} \Lambda^A D_{m} \Lambda^B \big) \notag \\
 &\, + \frac{i}{4} \Omega_{mna} \big( (\Sigma_a )^{AB}
 \bar{\Lambda}_A \bar{\sigma}^{mn} \bar{\Lambda}_B 
+ (\bar{\Sigma}_a )_{AB} \Lambda^A \sigma^{mn} \Lambda^B \big) \notag \\
 &\, + \frac{1}{2} (\Sigma_a )^{AB} \bar{\Lambda}_A \bar{\Lambda}_{D}
 (\mathcal{A}_a )^{D}{}_{B} - \frac{1}{2} (\bar{\Sigma}_a )_{AB}
 \Lambda^A (\mathcal{A}_a )^{B}{}_{D} \Lambda^{D} \Big].
\label{eq:on-shell_action}
\end{align}
Here $F_{mn} = \partial_{m} A_{n} - \partial_{n} A_{m} + i
[A_{m}, A_{n}]$ is the gauge field strength, $D_{m} * =
\partial_{m} * + i [A_{m}, * ]$ is the gauge covariant
derivative. 
$g=g_{10}V_{6}^{-\frac{1}{2}}$ is the coupling constant in four dimensions, 
where $V_{6}$ is the volume of the six-dimensional torus. 
$\sigma^m \ (\bar{\sigma}^m)$, $\Sigma_a \ (\bar{\Sigma}_a)$ are the
Dirac matrices in four and six dimensions, whose explicit forms are found in 
appendix \ref{sc:appA}. 
$G_{ma}, H_{ab}$ are defined by 
\begin{align}
G_{ma} =& \  D_m \varphi_a + \Omega_a^n F_{nm}, \notag \\
H_{ab} =& \ i [\varphi_a, \varphi_b] - \Omega_a^m D_m \varphi_b +
 \Omega^m_b D_m \varphi_a + \Omega_a^m \Omega_b^n F_{mn} - 
T_{ab} {}^c \varphi_c.
\label{eq:GHT}
\end{align}
The torsion is identified with the constant $SU(4)_I$
R-symmetry Wilson line gauge field $(\mathcal{A}_a)^A {}_B$ by the
following relation \cite{Ito:2012hs}:
\begin{align}
T_{ab} {}^c =& \ 
\frac{1}{2}
\big( (\Sigma_b \bar{\Sigma}_c
 )^{A}{}_{B} 
(\mathcal{A}_a )^{B}{}_{A} - (\Sigma_a \bar{\Sigma}_c )^{A}{}_{B} 
(\mathcal{A}_b )^{B}{}_{A}
 \big).
\label{torsion}
\end{align}
The four-dimensional supersymmetry transformation is obtained by the
dimensional reduction of \eqref{SUSYtransf}:
\begin{align}
\delta A_m =& \ - \xi^A \sigma_m \bar{\Lambda}_A - \bar{\zeta}^A
 \bar{\sigma}_m \Lambda_A,
\notag \\
\delta \Lambda^A =& \ \sigma^{mn} \zeta^A F_{mn} + i(\Sigma_a)^{AB} \sigma^m \bar{\zeta}_B G_{ma} + (\Sigma_{ab})^A {}_B \zeta^B H_{ab}, 
\notag \\
\delta \bar{\Lambda}_A =& \ \bar{\sigma}^{mn} \bar{\zeta}_A
 F_{mn} + i (\bar{\Sigma}_a)_{AB} \bar{\sigma}^m \bar{\zeta}^B
 G_{ma} + (\bar{\Sigma}_{ab})_A {}^B \bar{\zeta}_B H_{ab},
\notag \\
\delta \varphi_a =& \ i \zeta^A (\bar{\Sigma}_a)_{AB} \Lambda^B - i
 \bar{\zeta}_A (\Sigma_a)^{AB} \bar{\Lambda}_B
- \Omega^m_a \bar{\zeta}_A \bar{\sigma}_m \Lambda^A 
+ \Omega^m_a \zeta^A \sigma_m \bar{\Lambda}_A,
\label{eq:4DSUSY}
\end{align}
where 
$\Sigma_{ab}$, $\bar{\Sigma}_{ab}$ are the six-dimensional Lorentz
generators defined in appendix A.
The constant supersymmetry parameters 
$\zeta = (\zeta_{\alpha}{}^A, \bar{\zeta}^{\dot{\alpha}}{}_A)$ satisfy the parallel spinor
conditions in four dimensions:
\begin{align}
& \Omega_{mna} (\bar{\sigma}^{mn})^{\dot{\alpha}} {}_{\dot{\beta}}
 \bar{\zeta}^{\dot{\beta}} {}_A + 2 i (\mathcal{A}_a)_A {}^B
 \bar{\zeta}^{\dot{\alpha}} {}_B = 0, \notag \\
& \Omega_{mna} (\sigma^{mn})_{\alpha} {}^{\beta} \zeta_{\beta} {}^A + 2 i
 (\mathcal{A}_a)^A {}_B \zeta_{\alpha} {}^B = 0.
\label{parallel4D}
\end{align}
In \eqref{parallel4D}, 
the first terms represent the four-dimensional rotation of the spinor by the parameter
$\Omega_{mna}$.
The second terms represent the six-dimensional rotation by the parameter $(\mathcal{A}_a)^A {}_B$.
It is convenient to introduce the topological twist to see the cancellation between the four- and six-dimensional rotations.
The topological twists in $\mathcal{N} = 4$ super Yang-Mills theory are
classified into three types \cite{Yamron}.
They are called the half twist \cite{Witten:1988ze}, the Vafa-Witten twist \cite{Vafa:1994tf}
and the Marcus twist \cite{Marcus:1995mq,Kapustin:2006pk}.
We note that supersymmetry requires 
the additional conditions for $\Omega_{mna}$ and $(\mathcal{A}_a)^A {}_B$ 
\cite{Ito:2012hs}, where we find solutions of $\Omega_{mna}$ and
$(\mathcal{A}_a)^A {}_B$ for which $\zeta_{\alpha}{}^A, \bar{\zeta}^{\dot{\alpha}}{}_A$ satisfy \eqref{parallel4D}
and a part of supercharges is conserved.

Although a part of supersymmetry is preserved in each twisted theory, the 
four-dimensional translational symmetry is broken by the
$\Omega$-background \eqref{eq:Omega_matrix}.
In \cite{Ito:2012hs}, we have shown that the supersymmetries are
enhanced in the Nekrasov-Shatashvili (NS) limit 
$\epsilon^1_{a} \to 0$ (or $\epsilon^2_{a} \to 0$) 
\cite{Nekrasov:2009rc}, where the translational symmetry in a two-dimensional subspace is recovered.
In the following subsections, we summarize the solutions to the
conditions and the enhanced supersymmetries in the three types of the topological twists.

\subsection{Topological twist and supersymmetry in the NS limit}
The topological twist is defined by an embedding of the Lorentz group $SO(4) \simeq SU(2)_L \times
SU(2)_R$ into the subgroups of the $SU(4)_I$ R-symmetry group.
We take the $SU(2)_{L'} \times SU(2)_{R'}$ subgroup of $SU(4)_I$, 
where the $SU(4)_I$ index $A=1,2,3,4$ is decomposed into $A' = 1,2$ and $\hat{A} =
3,4$. Here $A'$ and $\hat{A}$ are indices for the two-dimensional representations of 
$SU(2)_{R'}$ and $SU(2)_{L'}$, respectively. 

\paragraph{Half twist}
In the half twist, $SU(2)_R$ subgroup of the Lorentz group $SO(4)$ is
replaced by the diagonal subgroup of $SU(2)_{R'} \times SU(2)_R$.
Then the new Lorentz group becomes $SU(2)_L \times [SU(2)_{R'} \times SU(2)_{R}]_{\mathrm{diag}}$, where the subscript ``diag'' stands for the
diagonal subgroup.
The spinor index $\dot{\alpha}$
is identified with 
the R-symmetry index $A'$
in the half twist.
We define the vector, the scalar and the anti-self-dual tensor
supercharges $Q_{m}$, $\bar{Q}$, $\bar{Q}_{mn}$ by
\begin{align}
\bar{Q} = \delta_{A'} {}^{\dot{\alpha}} \bar{Q}_{\dot{\alpha}} {}^{A'},
 \qquad 
\bar{Q}_{mn} = - (\bar{\sigma}_{mn})_{A'} {}^{\dot{\alpha}}
 \bar{Q}_{\dot{\alpha}} {}^{A'},
\qquad 
Q_m = - \varepsilon^{\alpha \beta} (\sigma_m)_{\beta A'} Q_\alpha{}^{A'}
,
\label{eq:half_fermion_decomp}
\end{align}
where $Q_{\alpha A}$, $\bar{Q}_{\dot{\alpha}}{}^A$ are the supercharges associated
with the supersymmetry transformation \eqref{eq:4DSUSY}.
The background $\Omega_{mna}$ and $(\mathcal{A}_a)^A {}_B$
such that $\bar{Q}$ and $\bar{Q}_{12}$ are conserved \cite{Ito:2012hs} is 
\begin{align}
& \Omega_{mna} 
= 
\left(
\begin{array}{cccc}
0 & \epsilon^1_{a} & 0 & 0 \\
- \epsilon^1_{a} & 0 & 0 & 0 \\
0 & 0 & 0 & - \epsilon^2_{a} \\
0 & 0 & \epsilon^2_{a} & 0
\end{array}
\right), \quad 
(\mathcal{A}_{a})^A {}_B = 
\left(
\begin{array}{cc}
\frac{1}{2} (\epsilon_{a}^1 + \epsilon_{a}^2) \tau^3 & 0 \\
0 & m_{a} \tau^3
\end{array}
\right), \quad (a=1,2), 
 \notag \\
& 
\Omega_{mna} = (\mathcal{A}_{a})^A {}_B = 0, 
\quad (a = 3,4,5,6),
\label{eq:half_SUSY_condition}
\end{align}
where $m_{a} \ (a=1,2)$ are real parameters.
These parameters are identified with the mass 
of the adjoint hypermultiplet in the $\mathcal{N} = 2^{*}$
theory \cite{Nekrasov:2003rj, Ito:2011cr}.
The theory has $\mathcal{N}=(0,2)$ supersymmetry for the
background \eqref{eq:half_SUSY_condition}.
Here the notation $\mathcal{N} = (m,n)$ means that the theory has $m$ chiral,
$n$ anti-chiral supercharges. 

In the NS limit\footnote{
In this paper, we usually consider the limit by $\epsilon^2_{a} \to
0$, where the translational symmetry in the $(x^3, x^4)$-plane is recovered.
}, two components of the vector supercharges 
$Q_3, Q_4$ are conserved in addition to the scalar and the tensor supercharges $\bar{Q}$, $\bar{Q}_{12}$. 
In the ordinary basis, the conserved supercharges in the NS limit
correspond to $(Q_{11},Q_{22})$ in $Q_{\alpha A}$ and $(\bar{Q}_{\dot{1}} {}^1, \bar{Q}_{\dot{2}} {}^2)$ in $\bar{Q}_{\dot{\alpha}}
{}^{A}$.
Therefore the supersymmetry is enhanced to $\mathcal{N} = (2,2)$.

\paragraph{Vafa-Witten twist}
In the Vafa-Witten twist, 
the new Lorentz group is $SU(2)_L \times [SU(2)_{L'} \times
SU(2)_{R'} \times SU(2)_R]_{\mathrm{diag}}$.
The spinor index $\dot{\alpha}$ is identified with 
the R-symmetry indices $A'$ and $\hat{A}$.
We define the two scalars $\bar{Q}$, $\hat{\bar{Q}}$, the two
vectors $Q_{m}$, $\hat{Q}_{m}$ and the two anti-self-dual tensor supercharges
$\bar{Q}_{mn}$, $\hat{\bar{Q}}_{mn}$ by 
\begin{align}
& Q_m = - \varepsilon^{\alpha \beta} (\sigma_m)_{\beta A'} Q_\alpha{}^{A'}
, \qquad 
\hat{Q}_m = - \varepsilon^{\alpha \beta} (\sigma_m)_{\beta \hat{A}} Q_\alpha{}^{\hat{A}}
, 
\notag \\
& \bar{Q} = \delta_{A'} {}^{\dot{\alpha}} \bar{Q}_{\dot{\alpha}}
 {}^{A'}, \qquad 
\hat{\bar{Q}} = \delta_{\hat{A}} {}^{\dot{\alpha}} \bar{Q}_{\dot{\alpha}}
 {}^{\hat{A}},
\notag \\
& \bar{Q}_{mn} = - (\bar{\sigma}_{mn})_{A'} {}^{\dot{\alpha}}
 \bar{Q}_{\dot{\alpha}} {}^{A'}, \qquad 
\hat{\bar{Q}}_{mn} = - (\bar{\sigma}_{mn})_{\hat{A}} {}^{\dot{\alpha}}
 \bar{Q}_{\dot{\alpha}} {}^{\hat{A}}.
\label{eq:VW_fermion_decomp}
\end{align}
The background $\Omega_{mna}$ and $(\mathcal{A}_a)^A {}_B$ 
such that $\bar{Q}$, $\bar{Q}_{12}$, $\hat{\bar{Q}}$ and $\hat{\bar{Q}}_{12}$ are conserved is
\begin{align}
& \Omega_{mna} 
= 
\left(
\begin{array}{cccc}
0 & \epsilon_{a}^1 & 0 & 0 \\
- \epsilon_{a}^1 & 0 & 0 & 0 \\
0 & 0 & 0 & - \epsilon_{a}^2 \\
0 & 0 & \epsilon_{a}^2 & 0
\end{array}
\right), \quad 
(\mathcal{A}_{a})^A {}_B 
= 
\left(
\begin{array}{cc}
\frac{1}{2} (\epsilon^1_{a} + \epsilon^2_{a}) \tau^3
 & 0 \\
0 & 
\frac{1}{2} (\epsilon^1_{a} + \epsilon^2_{a}) \tau^3
\end{array}
\right), 
\notag \\
& (a = 1,2,5,6), 
\notag \\
& \Omega_{mna} = (\mathcal{A}_a)^A {}_B = 0, \quad (a=3,4).
\label{eq:VW_SUSY_condition}
\end{align}
The theory has $\mathcal{N}
= (0,4)$ supersymmetry for the background \eqref{eq:VW_SUSY_condition}.
In the NS limit,
four vector supercharges $Q_3, Q_4, \hat{Q}_3, \hat{Q}_4$ 
are conserved in addition to $\bar{Q}, \hat{\bar{Q}}, \bar{Q}_{12},
\hat{\bar{Q}}_{12}$.
In the ordinary basis, these conserved supercharges correspond to 
$(Q_{11}, Q_{22}, Q_{13},Q_{24})$ in $Q_{\alpha A}$ and 
$(\bar{Q}_{\dot{1}} {}^{1},
\bar{Q}_{\dot{2}} {}^{2}, \bar{Q}_{\dot{1}} {}^{3}, \bar{Q}_{\dot{2}} {}^{4})$ in $\bar{Q}_{\dot{\alpha}} {}^{A}$.
Therefore the supersymmetry is enhanced to $\mathcal{N} = (4,4)$.

\paragraph{Marcus twist}
In the Marcus twist, 
the new Lorentz group is $[SU(2)_{L'} \times
SU(2)_{L}]_{\mathrm{diag}} \times [SU(2)_{R'} \times
SU(2)_{R}]_{\mathrm{diag}}$, where 
the spinor indices $\alpha$ and $\dot{\alpha}$
are identified with the R-symmetry indices $\hat{A}$ and $A'$, respectively.
We define the two scalars $Q$, $\bar{Q}$,
the two vectors $Q_{m}$, $\bar{Q}_{m}$ and the two tensor
supercharges $Q_{mn}$, $\bar{Q}_{mn}$ by 
\begin{align}
& Q = \delta^{\alpha} {}_{\hat{A}} Q_{\alpha} {}^{\hat{A}}, \qquad 
Q_m = - \varepsilon^{\alpha \beta} (\sigma_m)_{\beta A'} Q_\alpha{}^{A'}
, \qquad 
Q_{mn} = - (\sigma_{mn})^{\hat{A}} {}_{\alpha} Q^{\alpha} {}_{\hat{A}}, 
\notag \\
& \bar{Q} =  \delta_{\dot{\alpha}} {}^{A'} \bar{Q}^{\dot{\alpha}}
 {}_{A'}, \qquad 
\bar{Q}_m = - \varepsilon^{\dot{\alpha} \dot{\beta}}
 (\bar{\sigma}_m)_{\dot{\beta} \hat{A}} \bar{Q}_{\dot{\alpha}}
 {}^{\hat{A}},  
\qquad 
\bar{Q}_{mn} = - (\bar{\sigma}_{mn})_{A'} {}^{\dot{\alpha}}
 \bar{Q}_{\dot{\alpha}} {}^{A'}.
\label{eq:Marcus_fermion_decomp}
\end{align}
The background $\Omega_{mna}$ and $(\mathcal{A}_a)^A {}_B$  
such that the theory has $\mathcal{N}=(2,2)$ supersymmetry is
\begin{align}
& \Omega_{mna} 
= 
\left(
\begin{array}{cccc}
0 & \epsilon_{a}^1 & 0 & 0 \\
- \epsilon_{a}^1 & 0 & 0 & 0 \\
0 & 0 & 0 & - \epsilon_{a}^2 \\
0 & 0 & \epsilon_{a}^2 & 0
\end{array}
\right), \quad 
(\mathcal{A}_{a})^{A} {}_{B} 
= 
\left(
\begin{array}{cc}
\frac{1}{2} (\epsilon_{a}^1 + \epsilon_{a}^2) \tau^3
 & 0 \\
0 & 
\pm\frac{1}{2} (\epsilon_{a}^1 - \epsilon_{a}^2) \tau^3 
\end{array}
\right),
\notag \\
& (a = 1,2), 
\notag \\
& \Omega_{mna} = (\mathcal{A}_a)^{A} {}_B = 0, \quad 
(a = 3,4,5,6).
\label{eq:Marcus_SUSY_condition}
\end{align}
For the background with the minus sign in the lower-right block in $(\mathcal{A}_a)^A {}_B$, 
the conserved supercharges are
$Q$, $Q_{12}$, $\bar{Q}$ and $\bar{Q}_{12}$ \cite{Ito:2012hs}.
If we choose the plus sign, $Q_{13}$, $Q_{14}$ , $\bar{Q}$ and $\bar{Q}_{12}$  are conserved. 
But 
it can be shown
that the theory
defined in the background \eqref{eq:Marcus_SUSY_condition} with the plus sign
is the same as the one with the minus sign by suitable field redefinition.

In the NS limit, for the minus sign in the lower-right block in $(\mathcal{A}_a)^A {}_B$, four vector supercharges $Q_1, Q_2, \bar{Q}_1,
\bar{Q}_2$ are conserved in addition to $Q,
\bar{Q}, Q_{12}, \bar{Q}_{12}$.
In the ordinary basis, these supercharges correspond to 
$(Q_{11}, Q_{22}, Q_{14}, Q_{23})$ in $Q_{\alpha A}$ and
$(\bar{Q}_{\dot{1}} {}^{1}, \bar{Q}_{\dot{2}}
{}^{2}, \bar{Q}_{\dot{2}} {}^{3}, \bar{Q}_{\dot{1}} {}^{4})$ in $\bar{Q}_{\dot{\alpha}} {}^{A}$. 
The supersymmetry is enhanced to $\mathcal{N} = (4,4)$.
If we choose the plus sign, the conserved supercharges in the ordinary basis are 
$(Q_{11}, Q_{22}, Q_{13}, Q_{24})$ in $Q_{\alpha A}$ and  
$(\bar{Q}_{\dot{1}} {}^{1}, \bar{Q}_{\dot{2}}
{}^{2}, \bar{Q}_{\dot{1}} {}^{3}, \bar{Q}_{\dot{2}} {}^{4})$ in $\bar{Q}_{\dot{\alpha}} {}^{A}$, 
which are the same as 
the case of the Vafa-Witten twist. Moreover in the NS limit, 
the background is obtained from 
\eqref{eq:VW_SUSY_condition} by setting $\epsilon_{5}^1 = \epsilon_6^1 = 0$. 
Therefore the Marcus twist is regarded as the special case of 
the Vafa-Witten twist in the NS limit.

The conserved supercharges in the NS limit are summarized in
Table \ref{tb:SUSY_summary}.

\begin{table}[tb]
\begin{center}
\begin{tabular}{|c|c|}
\hline
Topological twist& Conserved supercharges $Q_{\alpha A}$, $\bar{Q}_{\dot{\alpha}}{}^A$ \\
\hline \hline
Half twist & 
$(Q_{11}, Q_{22})$, $(\bar{Q}_{\dot{1}} {}^1, \bar{Q}_{\dot{2}} {}^2)$
 \\
\hline
Vafa-Witten twist  & 
\multirow{2}{*}{
$(Q_{11}, Q_{22}, Q_{13}, Q_{24})$ , 
$(\bar{Q}_{\dot{1}} {}^{1}, \bar{Q}_{\dot{2}} {}^{2}, \bar{Q}_{\dot{1}} {}^{3}, 
\bar{Q}_{\dot{2}} {}^{4})$
}
 \\
\cline{1-1}
Marcus twist $(+)$ & 
 \\
\hline
Marcus twist $(-)$& 
$(Q_{11}, Q_{22}, Q_{14}, Q_{23})$ , 
$(\bar{Q}_{\dot{1}} {}^{1}, \bar{Q}_{\dot{2}} {}^{2}, 
\bar{Q}_{\dot{2}} {}^{3}, \bar{Q}_{\dot{1}} {}^{4})$ 
 \\
\hline
\end{tabular}
\caption{Conserved supercharges in the NS limit for topological twists.
The signs $(+)$ and $(-)$ in the Marcus twist correspond to the choice of 
the sign of $(\mathcal{A}_{a})^{\hat{A}}{}_{\hat{B}}$
in \eqref{eq:Marcus_SUSY_condition}. 
}
\label{tb:SUSY_summary}
\end{center}
\end{table}

\section{Central charges and BPS bounds in the NS limit}
We now study the supersymmetry algebra for $\Omega$-deformed
$\mathcal{N}=4$ super Yang-Mills theory in the NS limit
$\epsilon^{2}_{a}\to 0$ such that the translational symmetry in the $(x^3,x^4)$-plane is recovered.
We also perform the inverse Wick rotation 
$x^4 =  i x^0$ and consider the theory in the spacetime with the Minkowski 
signature, which implies that the energy and the third component of the momentum 
are well-defined and conserved. 
In this section we will calculate the central charges of the algebras and the BPS bounds for the mass.

We first discuss the dimensional reduction of the supersymmetry algebra 
\eqref{SUSYalg10D} in the $\Omega$-background.
The supersymmetry generator
$\bar{\zeta}\mathcal{Q}$ in ten dimensions is expressed in terms of the four-dimensional spinors $\zeta$, $\bar{\zeta}$, $Q$ and $\bar{Q}$ as 
\begin{gather}
\bar{\zeta}\mathcal{Q}=
\zeta^{\alpha A}Q_{\alpha A}
+\bar{Q}_{\dot{\alpha}}{}^{A}\bar{\zeta}^{\dot{\alpha}}{}_{A}. 
\end{gather}
The ten-dimensional momentum $P_{M}$ ($M=0,1,2,3,5,\ldots,10$) is decomposed into the four-dimensional momentum $P_{m}$ ($m=0,1,2,3$) and the six-dimensional momentum $P_{a+4}$ ($a=1,\ldots,6$).
The central charge $\mathcal{Z}_{MNPQR}$ is decomposed into $\mathcal{Z}_{a^\prime b^\prime c^\prime d^\prime e^\prime}$, $\mathcal{Z}_{ma^\prime b^\prime c^\prime d^\prime}$, $\mathcal{Z}_{mna^\prime b^\prime c^\prime}$ and $\mathcal{Z}_{mnpa^\prime b^\prime}$ ($a^\prime,\ldots, e^\prime =5,\ldots,10$).
Substituting the $\Omega$-background 
\eqref{Omegabcgdmetric} and the torsion \eqref{torsion}
into \eqref{SUSYalg10D}, we obtain the supersymmetry algebra in four dimensions:
\begin{align}
&\bigl[\xi Q+\bar{Q}\bar{\xi},\,\zeta Q+\bar{Q}\bar{\zeta}\bigr]
\notag\\[2mm]
&=
2(\xi\sigma^{m}\bar{\zeta}+\bar{\xi}\bar{\sigma}^{m}\zeta)P_{m}
+2(\xi\bar{\Sigma}_{a}\zeta-\bar{\xi}\Sigma_{a}\bar{\zeta})
P_{a+4}
-2i(\xi\bar{\Sigma}_{a}\zeta+\bar{\xi}\Sigma_{a}\bar{\zeta})
\biggl(\frac{1}{5!}\varepsilon^{abcdef}Z_{bcdef}\biggr)
\notag\\[2mm]
&\qquad{}
-i(\xi\sigma^{i}\bar{\Sigma}_{abcd}\bar{\zeta}
+\bar{\xi}\bar{\sigma}^{i}\Sigma_{abcd}\zeta)Z_{i,abcd}
+2(\xi\sigma^{ij}\bar{\Sigma}_{abc}\zeta
+\bar{\xi}\bar{\sigma}^{ij}\Sigma_{abc}\bar{\zeta})Z_{ij,abc}
\notag\\[2mm]
&\qquad{}
+\frac{i}{4}\varepsilon^{ijk}(\xi\sigma^0\bar{\Sigma}_{ab}\bar{\zeta}
-\bar{\xi}\bar{\sigma}^0\Sigma_{ab}\zeta)Z_{ijk,ab},
\label{SUSYalg4D}
\end{align}
where $Z_{abcde}\equiv \mathcal{Z}_{(a+4)\cdots (e+4)}$, $Z_{i,abcd} \equiv \mathcal{Z}_{i(a+4)\cdots (d+4)}$, etc. 
The indices $i,j,k=1,2,3$ denote the spatial components.
The symbols
$\varepsilon^{abcdef}$ and $\varepsilon^{ijk}$ 
represent
the totally antisymmetric 
tensors
in six and three dimensions, respectively. 
The matrices $\Sigma_{a_{1}a_{2}\cdots a_{n}}$,
$\bar{\Sigma}_{a_{1}a_{2}\cdots a_{n}}$ are the antisymmetrized products
of the Dirac matrices defined in appendix \ref{sc:appA}.

Now we calculate the central charges of the supersymmetry algebra \eqref{SUSYalg4D}.
The six-dimensional momentum $P_{a+4}$ is obtained by the spatial integral of the energy-momentum tensor \eqref{EMtensor10D}.
After the dimensional reduction, we have
\begin{align}
P_{a+4}
=
\int\!d^{3}x\,
\frac{1}{\kappa g^{2}}\mathrm{Tr}\biggl[F^{m0}G_{ma}+G^{0b}H_{ab}\biggr].
\label{eq:ele_charge}
\end{align}
Here the fermionic part is omitted since we are interested in 
the BPS configuration, in which the fermions are set to be zero. 
This momentum is interpreted as the electric charge \cite{Osborn:1979tq}.
Then we define the electric charge $q^{(e)}_{a}$ by
$
q^{(e)}_{a}
\equiv
P_{a+4}
$.

The central charge $Z_{bcdef}$ is related to the magnetic charge.
We define the magnetic charge $q^{(m)}_{a}$ by
the Hodge dual of $Z_{bcdef}$: 
\begin{gather}
q^{(m)}_{a}
=
\frac{1}{5!}\varepsilon^{abcdef}Z_{bcdef}
=
\int\!d^{3}x\,\frac{1}{\kappa g^{2}}
\mathrm{Tr}\biggl[\frac{1}{2}\varepsilon^{ijk}F_{ij}G_{ka}
\biggr].
\label{eq:mag_charge}
\end{gather}
The other charges $Z_{i,abcd}$, $Z_{ij,abc}$ 
and $Z_{ijk,ab}$ 
correspond to the charges of the BPS vortices, the BPS domain-walls and the space-filling BPS objects.
In the next subsections, we examine the $\Omega$-deformation of the central charges in \eqref{SUSYalg4D} 
for each twist. 

\subsection{Half twist}
\label{sc:half}

In the case of the half twist with the NS limit $\epsilon^{2}_{a}\to 0$, 
we have the conserved supercharges 
$Q_{11}$, $Q_{22}$, $\bar{Q}_{\dot{1}}{}^{1}$ and $\bar{Q}_{\dot{2}}{}^{2}$. 
The supersymmetry generator $\zeta^{\alpha A}Q_{\alpha A}+\bar{Q}_{\dot{\alpha}}{}^{A}\bar{\zeta}^{\dot{\alpha}}{}_{A}$ becomes
\begin{gather}
\zeta^{11}Q_{11}+\zeta^{22}Q_{22}
+\bar{Q}_{\dot{1}}{}^{1}\bar{\zeta}^{\dot{1}}{}_{1}
+\bar{Q}_{\dot{2}}{}^{2}\bar{\zeta}^{\dot{2}}{}_{2}.
\label{eq:susy_transf_half}
\end{gather}
Let us examine the right hand side in the supersymmetry algebra \eqref{SUSYalg4D}.
For the energy and the momentum, 
only $P^{0}$ and $P^{3}$ contribute to the algebra.
This is consistent with the recovery of the translational invariance for 
$x^{0}$- and $x^{3}$-directions under the NS limit. 
For the electric charge \eqref{eq:ele_charge} and the magnetic charge \eqref{eq:mag_charge}, 
in the present representation of the Dirac matrices,
the ones with indices $a=1,2$ only contribute.
These charges can be rewritten using the equations of motion
and the Bianchi identity as 
\begin{align}
q^{(e)}_{a}
&=
\int\!d^{3}x\,\frac{1}{\kappa g^{2}}\mathrm{Tr}\biggl[
D_{i}\bigl(F^{i0}\varphi_{a}\bigr)
+\sum_{b=1}^2\epsilon^{1}_{b}(x_{1}D_{2}-x_{2}D_{1})\bigl\{
(D_{0}\varphi_{b}-\epsilon^{1}_{b}x_{1}F_{20}+\epsilon^{1}_{b}x_{2}F_{10})
\varphi_{a}\bigr\}
\notag\\
&\qquad{}
+\frac{1}{2}(\epsilon^{1}_{a}+2m_{a})
(\varphi_{3}D_{0}\varphi_{4}-\varphi_{4}D_{0}\varphi_{3})
-\frac{1}{2}(\epsilon^{1}_{a}-2m_{a})
(\varphi_{5}D_{0}\varphi_{6}-\varphi_{6}D_{0}\varphi_{5})
\notag\\
&\qquad{}
+\epsilon^{1}_{a}(x_{1}\mathcal{T}^{0}{}_{2}-x_{2}\mathcal{T}^{0}{}_{1})
\biggr],
\label{ele_def}
\\[2mm]
q^{(m)}_{a}
&=
\int\!d^{3}x\,\frac{1}{\kappa g^{2}}
\mathrm{Tr}\biggl[\frac{1}{2}\varepsilon^{ijk}D_{k}(F_{ij}\varphi_{a})\biggr],
\qquad (a=1,2),
\label{mag}
\end{align}
where $\mathcal{T}^{m}{}_{n}$ is the bosonic part of the four-dimensional energy momentum tensor:
\begin{align}
\mathcal{T}^{m}{}_{n}
&=
\frac{1}{\kappa g^{2}}\mathrm{Tr}\biggl[
F^{mp}F_{np}+G^{ma}G_{na}
-\frac{1}{4}\delta^{m}{}_{n}\Bigl(F^{pq}F_{pq}+2G^{pa}G_{pa}+H^{ab}H_{ab}\Bigr)
\biggr].
\end{align}
The first term in the right hand side of \eqref{ele_def} is the electric charge density of the undeformed theory. 
The remaining terms in the first line 
take the form of total derivatives
and give the deformation of the electric charge.
The second line is the R-charge density. 
The third line is the angular momentum density in the $(x^{1},x^{2})$-plane. 
Setting
 $\varphi_3,\ldots,\varphi_6=0$
in \eqref{ele_def}
and imposing
$\varphi_2=0$ and $\epsilon^1_2=0$,
one can
reproduces the electric charge
of the $\Omega$-deformed $\mathcal{N}=2$ theory
\cite{Ito:2011ta}. 
The magnetic charge in \eqref{mag} is not deformed.

For the vortex charge $Z_{i,abcd}$, it is convenient to consider its Hodge dual, 
which is given by 
\begin{gather}
V_{ijab}
=
\frac{1}{4!}\varepsilon^{ijk}\varepsilon^{abcdef}Z_{k,cdef}
=
\int\!d^{3}x\,\frac{1}{\kappa g^{2}}\mathrm{Tr}\Bigl[
F_{ij}H_{ab}-G_{ia}G_{jb}+G_{ja}G_{ib}
\Bigr].
\label{eq:vortex_charge}
\end{gather}
From the spinorial structure of the conserved supercharges, the components $V_{1212}$, $V_{1234}$ and $V_{1256}$ can contribute to the algebra \eqref{SUSYalg4D}.
These are computed as
\begin{align}
V_{ijab}
&=
\int\!d^{3}x\,\frac{1}{\kappa g^{2}}
\mathrm{Tr}\Bigl[iF_{ij}[\varphi_{a},\varphi_{b}]
-D_{i}\varphi_{a}D_{j}\varphi_{b}
+D_{j}\varphi_{a}D_{i}\varphi_{b}
\notag\\
&\qquad\qquad\qquad\qquad{}
-3\Omega^{k}_{a}F_{[ij}D_{k]}\varphi_{b}
+3\Omega^{k}_{b}F_{[ij}D_{k]}\varphi_{a}
\notag\\
&\qquad\qquad\qquad\qquad{}
+\Omega^{k}_{a}\Omega^{l}_{b}
\bigl(F_{ij}F_{kl}
-F_{ik}F_{jl}
+F_{il}F_{jk}\bigr)
\Bigr],
\label{calc_vortex}
\end{align}
where we introduced the notation of antisymmetrization of indices $A_{[i_1 \cdots i_n]}$ with the normalization $1/n!$.
The first line in the right hand side in \eqref{calc_vortex} becomes the total derivative. 
The corresponding surface term depends on the values of the field configuration at infinity, which satisfies the vacuum condition $F_{mn}=G_{ma}=H_{ab}=0$. 
In particular the values $\langle\varphi_{a}\rangle$ of the scalar fields at infinity satisfy the equation
\begin{gather}
i\bigl[\langle\varphi_{a}\rangle, \langle\varphi_{b}\rangle\bigr]
-T_{ab}{}^{c}\langle\varphi_{c}\rangle=0.
\end{gather}
Here the torsion $T_{ab}{}^{c}$ is written in terms of the deformation parameters and is given in appendix \ref{sc:appB}. 
In the case of the half twist, $\langle\varphi_{1}\rangle$ and $\langle\varphi_{2}\rangle$ can be nonzero, which
commute with each other, while $\langle\varphi_{3}\rangle$, $\langle\varphi_{4}\rangle$, $\langle\varphi_{5}\rangle$ and $\langle\varphi_{6}\rangle$ are zero.
Using this vacuum configuration we find that the surface term from the first line in \eqref{calc_vortex} vanishes.
The second and the third lines vanish also in the NS limit.
Therefore 
$V_{ijab}$ does not contribute
 to the algebra \eqref{SUSYalg4D}.
In a similar way, we can show that 
neither
$Z_{ij,abc}$ nor $Z_{ijk,ab}$
contributes to the algebra.
This implies that under the vacuum boundary condition the BPS vortices, 
the BPS domain walls and the space-filling BPS objects do not exist in this theory.
This is similar to the case of the undeformed $\mathcal{N}=2^\ast$ and $\mathcal{N}=4$ theories. 
For $\Omega$-deformed $\mathcal{N}=2$ super Yang-Mills theory, see \cite{Bulycheva:2012ct}.
We finally obtain the supersymmetry algebra 
\begin{align}
\{Q_{11}, \bar{Q}_{\dot{1}}{}^{1}\}&=2(P^{0}+P^{3}), 
&
\{Q_{22}, \bar{Q}_{\dot{2}}{}^{2}\}&=2(P^{0}-P^{3}),
\notag\\
\{Q_{11}, Q_{22}\}&=2Z, 
&
\{\bar{Q}_{\dot{1}}{}^{1}, \bar{Q}_{\dot{2}}{}^{2}\}&=2\bar{Z}, 
\label{Q_anticom}
\end{align}
and the other (anti-)commutators vanish. Here the central charge $Z$ is given by 
\begin{gather}
Z=(q^{(m)}_{1}+iq^{(m)}_{2})+i(q^{(e)}_{1}+iq^{(e)}_{2}). 
\label{eq:center_Z}
\end{gather}
From \eqref{Q_anticom} the BPS bound for the mass $M$ 
becomes
\begin{gather}
M\ge |Z|=\sqrt{\bigl(q_{1}^{(m)}-q_{2}^{(e)}\bigr)^{2}
+\bigl(q_{2}^{(m)}+q_{1}^{(e)}\bigr)^{2}}\ . 
\label{eq:BPSbound_half}
\end{gather}
The supercharges preserved by the BPS states are given by 
\begin{gather}
Q_{11}-e^{i\theta}\bar{Q}_{\dot{2}}{}^{2},\quad 
\bar{Q}_{\dot{1}}{}^{1}-e^{-i\theta}Q_{22}. 
\label{eq:preserved_Q_half}
\end{gather}
Here $\theta$ is the argument of $Z$ and satisfies
\begin{gather}
\tan\theta=\frac{q_2^{(e)} + q_1^{(m)}}{q_1^{(e)} - q_2^{(m)}}.
\label{eq:theta}
\end{gather}
Since two supercharges out of four are preserved, 
these states are the 1/2 BPS states.

\subsection{Vafa-Witten and Marcus twists}
In the case of the Vafa-Witten and the Marcus twists with the NS limit, we have the conserved 
supercharges
$Q_{11}$, $Q_{22}$, $Q_{13}$, $Q_{24}$, 
$\bar{Q}_{\dot{1}}{}^{1}$, $\bar{Q}_{\dot{2}}{}^{2}$, 
$\bar{Q}_{\dot{1}}{}^{3}$ and $\bar{Q}_{\dot{2}}{}^{4}$. 
The supersymmetry generator $\zeta^{\alpha A}Q_{\alpha A}
+\bar{Q}_{\dot{\alpha}}{}^{A}\bar{\zeta}^{\dot{\alpha}}{}_{A}$ is 
\begin{align}
\zeta^{11}Q_{11}+\zeta^{22}Q_{22}
+\zeta^{13}Q_{13}+\zeta^{24}Q_{24}
+\bar{Q}_{\dot{1}}{}^{1}\bar{\zeta}^{\dot{1}}{}_{1}
+\bar{Q}_{\dot{2}}{}^{2}\bar{\zeta}^{\dot{2}}{}_{2}
+\bar{Q}_{\dot{1}}{}^{3}\bar{\zeta}^{\dot{1}}{}_{3}
+\bar{Q}_{\dot{2}}{}^{4}\bar{\zeta}^{\dot{2}}{}_{4}.
\label{eq:susy_transf_VW}
\end{align}
From the vacuum conditions given in appendix \ref{sc:appB},
we can show that 
the supersymmetry algebra \eqref{SUSYalg4D} does not contain the central charge for the BPS vortices, the BPS domain walls and 
the space-filling BPS objects, as in the case of the half twist. 
The supersymmetry algebra reads
\begin{align}
\{Q_{11}, \bar{Q}_{\dot{1}}{}^{1}\}=
\{Q_{13}, \bar{Q}_{\dot{1}}{}^{3}\}&=2(P^{0}+P^{3}), 
&
\{Q_{22}, \bar{Q}_{\dot{2}}{}^{2}\}=
\{Q_{24}, \bar{Q}_{\dot{2}}{}^{4}\}&=2(P^{0}-P^{3}),
\notag\\
\{Q_{11}, Q_{22}\}&=
2i(q_1+iq_2),
&
\{\bar{Q}_{\dot{1}}{}^{1}, \bar{Q}_{\dot{2}}{}^{2}\}&=
-2i(\bar{q}_1-i\bar{q}_2),
\notag\\
\{Q_{13}, Q_{22}\}&=
2i(q_5+iq_6),
&
\{\bar{Q}_{\dot{1}}{}^{3}, \bar{Q}_{\dot{2}}{}^{2}\}&=
-2i(\bar{q}_5-i\bar{q}_6),
\notag\\
\{Q_{13}, Q_{24}\}&=
2i(q_1-iq_2),
&
\{\bar{Q}_{\dot{1}}{}^{3}, \bar{Q}_{\dot{2}}{}^{4}\}&=
-2i(\bar{q}_1+i\bar{q}_2),
\notag\\
\{Q_{11}, Q_{24}\}&=
-2i(q_5-iq_6), 
&
\{\bar{Q}_{\dot{1}}{}^{1}, \bar{Q}_{\dot{2}}{}^{4}\}&=
2i(\bar{q}_5+i\bar{q}_6),
\label{Q_anticom_VW}
\end{align}
where 
$q_a$ ($a=1,2,5,6$)
are the complexified electric-magnetic charges defined by
\begin{align}
q_a=q^{(e)}_a-iq^{(m)}_a.
\label{eq:center_VW}
\end{align}
Here the electric charges $q^{(e)}_{a}$ 
in \eqref{eq:ele_charge} and the magnetic charges $q^{(m)}_a$ in \eqref{eq:mag_charge} become
\begin{align}
q^{(e)}_{a}
&=
\int\!d^{3}x\,\frac{1}{\kappa g^{2}}\mathrm{Tr}\biggl[
D_{i}\bigl(F^{i0}\varphi_{a}\bigr)
+\sum_{b=1,2,5,6}\epsilon^{1}_{b}(x_{1}D_{2}-x_{2}D_{1})\bigl\{
(D_{0}\varphi_{b}-\epsilon^{1}_{b}x_{1}F_{20}+\epsilon^{1}_{b}x_{2}F_{10})
\varphi_{a}\bigr\}
\notag\\[2mm]
& \qquad\qquad{}
+\epsilon^{1}_{a}
\bigl(\varphi_{3}D_{0}\varphi_{4}-\varphi_{4}D_{0}\varphi_{3}
+x_{1}\mathcal{T}^{0}{}_{2}-x_{2}\mathcal{T}^{0}{}_{1}\bigr)
\biggr],
\label{ele_def_VW}
\\[2mm]
q^{(m)}_{a}
&=
\int\!d^{3}x\,\frac{1}{\kappa g^{2}}
\mathrm{Tr}\biggl[\frac{1}{2}\varepsilon^{ijk}D_{k}(F_{ij}\varphi_{a})\biggr],
\qquad (a=1,2,5,6).
\end{align}
From \eqref{Q_anticom_VW} the BPS bound is given by
\begin{align}
M & \ge 
\sqrt{|q^{(e)}|^{2}+|q^{(m)}|^{2}\pm 2|q^{(e)}||q^{(m)}|\sin\alpha},
\label{eq:BPSbound_VW}
\end{align}
where $|q^{(e)}|$ and $|q^{(m)}|$ are the lengths of the vectors 
$q^{(e)}=(q^{(e)}_{1},\,q^{(e)}_{2},\,q^{(e)}_{5},\,q^{(e)}_{6})$ and 
$q^{(m)}=(q^{(m)}_{1},\,q^{(m)}_{2},\,q^{(m)}_{5},\,q^{(m)}_{6})$, 
respectively. 
$\alpha$ is the angle between the vectors $q^{(e)}$ and $q^{(m)}$. 
The bound \eqref{eq:BPSbound_VW} has the same form as 
the undeformed theory
\cite{Osborn:1979tq, Fraser:1997nd}
except that the charges with the indices $a=3,4$ do not contribute.
This is because those charges vanish due to the vacuum conditions.
The number of supercharges preserved by the BPS states
depends on whether $\sin\alpha$ is zero or nonzero.

In the case of $\sin\alpha=0$, 
the supercharges preserved by the BPS states are
\begin{align}
Q_{11}-v_1\bar{Q}_{\dot{2}}{}^{2}+v_2\bar{Q}_{\dot{2}}{}^{4}, 
\quad 
Q_{13}-v_3\bar{Q}_{\dot{2}}{}^{2}-v_4\bar{Q}_{\dot{2}}{}^{4}, 
\label{eq:preserved_Q_VW2}
\end{align}
and their complex conjugates from the algebras \eqref{Q_anticom_VW}. 
Here the coefficients $v_{i}$ $(i=1,\ldots,4)$ are defined by
\begin{align}
v_{1}&=\frac{i(q_1+iq_2)}{\sqrt{|q^{(m)}|^2+|q^{(e)}|^2}}, & 
v_{2}&=\frac{i(q_5-iq_6)}{\sqrt{|q^{(m)}|^2+|q^{(e)}|^2}}, \notag \\
v_{3}&=\frac{i(q_5+iq_6)}{\sqrt{|q^{(m)}|^2+|q^{(e)}|^2}}, &
v_{4}&=\frac{i(q_1-iq_2)}{\sqrt{|q^{(m)}|^2+|q^{(e)}|^2}}. 
\label{eq:v_parameter}
\end{align}
Since four supercharges out of eight are preserved, 
these states are the 1/2 BPS states. 

For nonzero $\sin\alpha$, 
the supercharges preserved by the BPS states are 
\begin{gather}
w_1Q_{11}+w_2Q_{13}
+w_3\bar{Q}_{\dot{2}}{}^{2}
+w_4\bar{Q}_{\dot{2}}{}^{4},
\label{eq:preserved_Q_VW1}
\end{gather}
and its complex conjugate.
Here the coefficients $w_{i}$ 
$(i=1,\ldots,4)$ are defined by
\begin{align}
\left(
\begin{array}{c}
w_1 \\
w_2 \\
w_3 \\
w_4 \\
\end{array}
\right)
=
\frac{1}{|A|^2+|B|^2+|C|^2+|D|^2}
\left(
\begin{array}{c}
A \\
B \\
C \\
D \\
\end{array}
\right),
\label{eq:w_parameter}
\end{align}
where
\begin{align}
A &  = \frac{1}{2}[(q_1-iq_2)(\bar{q}_5+i\bar{q}_6)-(q_5+iq_6)(\bar{q}_1-i\bar{q}_2)], \notag \\
B & = -\frac{i}{2}(q_1\bar{q}_2-q_2\bar{q}_1-q_5\bar{q}_6+q_6\bar{q}_5) \mp |q^{(e)}||q^{(m)}|\sin\alpha, \notag \\
C & = -i\frac{(q_1+iq_2)A + (q_5+iq_6)B}{\bigl(|q^{(e)}|^{2}+|q^{(m)}|^{2} \pm 2|q^{(e)}||q^{(m)}|\sin\alpha
\bigr)^{1/2}}, \notag \\
D & = -i\frac{(q_5-iq_6)A - (q_1-iq_2)B}{\bigl(|q^{(e)}|^{2}+|q^{(m)}|^{2} \pm 2|q^{(e)}||q^{(m)}|\sin\alpha
\bigr)^{1/2}}.
\end{align}
Since two supercharges out of eight are preserved, 
these states are the 1/4 BPS states. 
Note that $|q^{(e)}||q^{(m)}|\sin\alpha$ is written in terms of the complexified charges \eqref{eq:center_VW} as
\begin{align}
 |q^{(e)}|^2|q^{(m)}|^2\sin^2\alpha = 
\frac{1}{8}
\sum_{
\begin{subarray}{c}a,b=1,2,5,6\end{subarray}
}
|q_a\bar{q}_b-q_b\bar{q}_a|^2
.
\end{align}
The limit $\alpha\to0$ is equivalent to $|q_a\bar{q}_b-q_b\bar{q}_a|\to0$ for any $a$, $b$.
Although $A$, $B$, $C$ and $D$ become zero in this limit, we can take the limit $\alpha\to0$ keeping either $w_{3,4}/w_1$ or $w_{3,4}/w_2$ fixed. 
We then
obtain the parameters $v_i$ for the 1/2 BPS states.

\section{$\Omega$-deformed BPS equations}
In this section
we study the BPS equations deformed in the $\Omega$-background in the NS limit.
These equations can be obtained by setting the supersymmetry transformations \eqref{eq:4DSUSY}
of the fermions to be zero.
In the undeformed theory, we have the BPS equations for the 1/2 BPS dyons and the 1/4 BPS dyons \cite{Fraser:1997nd, Weinberg:2006rq}. 
The 1/2 BPS dyon solutions are preserved by the supersymmetry transformations with the parameters satisfying the conditions
\begin{align}
\zeta_\alpha{}^{A^\prime} = -e^{-i\theta}\delta_{\alpha \dot{\alpha}}\varepsilon^{A^\prime B^\prime}\bar{\zeta}^{\dot{\alpha}}{}_{B^\prime}, \quad \zeta_\alpha{}^{\hat{A}} = -e^{-i\theta}\delta_{\alpha\dot{\alpha}}\varepsilon^{\hat{A}\hat{B}}\bar{\zeta}^{\dot{\alpha}}{}_{\hat{B}}.
\label{half_BPS}
\end{align}
Here we decompose the R-symmetry index $A$ into $A^\prime=1,2$ and $\hat{A}=3,4$.
$\theta$ is a real parameter.
Substituting \eqref{half_BPS} to the vanishing
condition for the supersymmetry transformations \eqref{eq:4DSUSY} 
of the fermions without the deformation, we obtain the 1/2 BPS dyon equations:
\begin{align}
 & E_i+\sin\theta D_i\varphi_1=0, \notag \\
& B_i+\cos\theta D_i\varphi_1=0, \notag \\
& D_0\varphi_1=0, \notag \\
& D_m\varphi_{c}=0, \quad (c=2,\ldots,6), \notag \\
& [\varphi_a, \varphi_b]=0, \quad (a,b=1,\ldots,6).
\label{halfdyonequndef}
\end{align}
Here the electric field $E_i$ and the magnetic field $B_i$ are introduced by $E_i=F_{0i}$, $B_i=\frac{1}{2}\varepsilon_{ijk}F_{jk}$.
The 1/4 BPS dyons are preserved by the supersymmetry with the parameters satisfying 
\begin{align}
\zeta_\alpha{}^{A^\prime} = -e^{-i\theta}\delta_{\alpha \dot{\alpha}}\varepsilon^{A^\prime B^\prime}\bar{\zeta}^{\dot{\alpha}}{}_{B^\prime},
\quad \zeta_\alpha{}^{\hat{A}}=\bar{\zeta}^{\dot{\alpha}}{}_{\hat{A}}=0.
\label{quarter_BPS}
\end{align}
In a similar way we obtain the 1/4 BPS dyon equations.
The parameter $\theta$ disappears in the BPS equations by defining
\begin{align}
\phi_1=\varphi_1\cos\theta+\varphi_2\sin\theta, \quad \phi_2=-\varphi_1\sin\theta+\varphi_2\cos\theta.
\label{scalar_rotation}
\end{align}
The 1/4 BPS dyon equations become
\begin{align}
& E_i-D_i\phi_2=0, \notag \\
& B_i+D_i\phi_1=0, \notag \\
& D_0\phi_1-i[\phi_1, \phi_2]=0, \notag \\
& D_0\phi_2=0, \notag \\
& D_i\varphi_a=0, \notag  \\
& D_0\varphi_a+i[\phi_2, \varphi_a]=0, \notag \\
& [\phi_1, \varphi_a]=0, \quad (a=3,4,5,6), \notag \\
& [\varphi_3, \varphi_6]-[\varphi_4,\varphi_5]=[\varphi_3, \varphi_5]+[\varphi_4, \varphi_6]=[\varphi_3, \varphi_4]-[\varphi_5, \varphi_6]=0.
\label{quarterdyonequndef}
\end{align}
Note that the 1/2 BPS equations are consistent with the equations of motion.
But for the 1/4 BPS equations, we have to impose the Gauss' law constraints:
\begin{align}
 & D_iE_i +i\sum_{a=1}^6[\varphi_a, D_0\varphi_a]=0,
\end{align}
 so that they are consistent with the equations of motion \cite{Lee:1998nv}.

\subsection{1/2 BPS equations for the half twist}
First we consider the 1/2 BPS dyon equations in the theory with the half twist.
The BPS states in this theory preserve the supercharges 
\eqref{eq:preserved_Q_half}, 
which are parametrized by $\zeta$ satisfying
\begin{gather}
\zeta^{11}=-e^{-i\theta}\bar{\zeta}^{\dot{2}}{}_{2}, \quad
\zeta^{22}=-e^{-i\theta}\bar{\zeta}^{\dot{1}}{}_{1}.
\label{eq:BPS_condition_half}
\end{gather}
Requiring the supersymmetry transformation \eqref{eq:4DSUSY} of the fermions $\delta \Lambda$ and $\delta \bar{\Lambda}$ to be zero for \eqref{eq:BPS_condition_half}, 
we obtain the 1/2 BPS dyon equations. 
As in the undeformed theory, 
it is convenient to remove the explicit $\theta$-dependence
from the BPS equations.
In addition to \eqref{scalar_rotation} we define
\begin{align}
\varepsilon_1^1& \equiv \cos\theta\,\epsilon_1^1+\sin\theta\,\epsilon_2^1,
& 
\varepsilon_1^2& \equiv -\sin\theta\,\epsilon_1^1+\cos\theta\,\epsilon_2^1,
\notag\\
\mu_1& \equiv \cos\theta\,m_1+\sin\theta\,m_2, & 
\mu_2& \equiv -\sin\theta\,m_1+\cos\theta\,m_2, 
\notag \\
\omega^{mn}_1 & \equiv \cos\theta\,\Omega^{mn}_1 + \sin\theta\,\Omega^{mn}_2, 
&
\omega^{mn}_2 & \equiv -\sin\theta\,\Omega^{mn}_1 + \cos\theta\,\Omega^{mn}_2. 
\label{para-rot}
\end{align}
The 1/2 BPS dyon equations become
\begin{align}
& E_i -(D_i\phi_2+\omega_2^mF_{mi})=0, \notag \\
& B_i +( D_i\phi_1+\omega^m_1F_{mi} )-i\delta_{3,i}\left([\varphi_3, \varphi_4] - [\varphi_5, \varphi_6]\right)=0, \notag \\
& D_0\phi_1+\omega^m_1F_{m0} -i[\phi_1, \phi_2]+\omega^m_1D_m\phi_2-\omega^m_2D_m\phi_1=0, \notag \\
& D_0\phi_2 +\omega^m_2F_{m0} = 0, \notag \\
& (D_1+iD_2)(\varphi_3-i\varphi_4)=0,\notag \\
& (D_1+iD_2)(\varphi_5+i\varphi_6)=0,\notag \\
& D_3\varphi_a+i(-1)^{f(a)}\biggl([\phi_1,\varphi_{f(a)}]+i\omega^m_1D_m\varphi_{f(a)}-(-1)^{f(a)}\frac{i}{2}(\varepsilon^1_1+(-1)^{g(a)}\mu_1)\varphi_a \biggr)=0,\notag \\
& D_0\varphi_a+i[\phi_2,\varphi_a]-\omega^m_2D_m\varphi_a+(-1)^{f(a)}\frac{1}{2}(\varepsilon_2^1+(-1)^{g(a)}\mu_2)\varphi_{f(a)}=0, \quad (a=3,4,5,6), \notag \\
& [\varphi_3, \varphi_6]-[\varphi_4,\varphi_5]=[\varphi_3, \varphi_5]+[\varphi_4, \varphi_6]=0,
\label{eq:BPSeqhalf}
\end{align}
where $f(a) \equiv a-(-1)^a$ and $g(a) \equiv [a/5]$ ($a=3,4,5,6$). 
Here $[x]$ denotes Gauss' symbol.
Their values are given by
\begin{align}
  f(3)&=4, \quad  f(4)=3 ,\quad f(5)=6,\quad  f(6)=5,  \notag \\
  g(3)&=0,\quad  g(4)=0,\quad  g(5)=1 ,\quad  g(6)=1. 
\label{eq:def_fandg}
\end{align}
We note that the supersymmetry condition \eqref{eq:BPS_condition_half} is included in that of \eqref{quarter_BPS}.
So we compare the 1/2 BPS dyon equations \eqref{eq:BPSeqhalf} with the undeformed 1/4 BPS dyon equations \eqref{quarterdyonequndef} rather than the 1/2 BPS dyon equations \eqref{halfdyonequndef}.
The equations for the electric and the magnetic fields are deformed. 
For the scalar fields, the equations with the derivative along the zeroth and the third directions are also deformed.
The equations for the scalar fields $\varphi_3,\ldots,\varphi_6$ with the derivative along the $x^1$- and $x^2$-directions are the holomorphic form.
On the other hand the equations of the antiholomorphic part 
do not appear because the corresponding supersymmetry is broken by the $\Omega$-deformation.
These holomorphic equations are the vortex-like equations.
Note that we have to impose the Gauss' law constraints in the NS limit:
\begin{align}
 & D_iE_i - \sum_{a=1,2}\biggl( D^m[\Omega_{ma}(D_0\varphi_a+\Omega^p_aF_{p0})] - i[\varphi_a, D_0\varphi_a+\Omega^p_aF_{p0}] \biggr) +i\sum_{a=3}^6[\varphi_a, D_0\varphi_a]=0,
\end{align}
 so that the BPS equations are consistent with the equations of motion, as in the case of the 1/4 BPS equations of the undeformed theory.

\subsection{1/2 BPS equations for the Vafa-Witten and the Marcus twists}
We next consider the 1/2 BPS dyon equations in the theory with the
Vafa-Witten and the Marcus twists.
The BPS states preserve the supercharges \eqref{eq:preserved_Q_VW2}. 
The parameters $\zeta$ and $\bar{\zeta}$ of the supersymmetry transformation \eqref{eq:susy_transf_VW} satisfy the conditions
\begin{align}
\zeta^{22}&=
-\bar{v}_{1}\bar{\zeta}^{\dot{1}}{}_{1}
-\bar{v}_{3}\bar{\zeta}^{\dot{1}}{}_{3}\,, & 
\zeta^{24}&=
\bar{v}_{2}\bar{\zeta}^{\dot{1}}{}_{1}
-\bar{v}_{4}\bar{\zeta}^{\dot{1}}{}_{3}\,,
\label{eq:BPS_cond_VW}
\end{align}
and their complex conjugates. 
As in the case of the half twist, we have the 1/2 BPS dyon equations.
But in order to obtain the BPS equations which have 
the nontrivial electric and magnetic fields, 
we need to impose the following conditions for $v_{i}$'s:
\begin{align}
 (v_1, v_2, v_3, v_4)= (v_1, 0, 0, \pm v_1), \quad \text{or} \quad (0, v_2, \pm v_2, 0).
\end{align}
These four cases are equivalent up to the field redefinition by the R-symmetry transformation. We consider the case $v_1=v_4$ and $v_2=v_3=0$, where $q_2=q_5=q_6=0$. Since we have $|v_{1}|=1$ from 
\eqref{eq:v_parameter}, we can introduce the angle $\theta$ by
\begin{align}
 v_1 = \frac{iq_1}{|q_1|} = e^{i\theta}.
\label{eq:theta1/2BPSVW}
\end{align}
In this case, the preserved supercharges are given by
\begin{gather}
Q_{11}-e^{i\theta}\bar{Q}_{\dot{2}}{}^{2}, 
\quad 
Q_{13}-e^{i\theta}\bar{Q}_{\dot{2}}{}^{4}, 
\label{preserved_1/2BPS}
\end{gather}
and their complex conjugates. 
These are included in the supercharges preserved by the undeformed 1/2 BPS equations.
The deformed 1/2 BPS dyon equations in the Vafa-Witten twist are
\begin{align}
& E_i+(D_i\varphi_1+\Omega^m_1F_{mi})\sin\theta =0, \notag \\
& B_i+(D_i\varphi_1+\Omega^m_1F_{m i})\cos\theta -i\delta_{3,i}[\varphi_3, \varphi_4] =0, \notag \\
& D_0\varphi_1+\Omega^m_1F_{m0} =0, \notag \\
& D_m\varphi_{a}+\Omega^n_{a}F_{nm}=0 , \quad (a=2,5,6), \notag \\
& (D_1+iD_2)(\varphi_3-i\varphi_4)=0, \notag \\
& D_3\varphi_b+i(-1)^{f(b)}\Bigl([\varphi_1,\varphi_{f(b)}]+i\Omega^m_1D_m\varphi_{f(b)}-i(-1)^{f(b)}\epsilon^1_1\varphi_b \Bigr)\cos\theta =0,\notag \\
& D_0\varphi_b+i\Bigl([\varphi_1,\varphi_b]+i\Omega^m_1D_m\varphi_{b}+i(-1)^{f(b)}\epsilon^1_1\varphi_{f(b)} \Bigr)\sin\theta =0,\quad (b=3,4),  \notag \\
& H_{cd}=0, \quad \bigl(1\leq c<d \leq 6, \ (c,d)\neq (1,3), (1,4), (3,4)\bigr),
\label{eq:halfBPSeqVW}
\end{align}
where the function $f(a)$ is given by \eqref{eq:def_fandg}.
Comparing the deformed 1/2 BPS equations \eqref{eq:halfBPSeqVW} with the undeformed 1/2 BPS equation \eqref{halfdyonequndef},
we find that the equations for the electric and the magnetic fields are deformed only by the deformation parameter $\epsilon_1^1$.
The other deformation parameters appear in the equations for the scalar fields.
For the scalar fields $\varphi_3$, $\varphi_4$ the equations along the $x^1$- and $x^2$-directions are the holomorphic forms, but the equations for the scalar fields $\varphi_5$, $\varphi_6$ are not holomorphic.
Note that 
\eqref{eq:halfBPSeqVW} is consistent with 
the equations of motion without imposing the Gauss' law constraints.

\subsection{1/4 BPS equations for the Vafa-Witten and the Marcus twists}

We finally consider the deformed 1/4 BPS dyon equations in the theory
with the Vafa-Witten and the Marcus twists. 
The BPS states preserve the supercharges \eqref{eq:preserved_Q_VW1}.
The parameters $\zeta$, $\bar{\zeta}$ of the supersymmetry transformations \eqref{eq:susy_transf_VW} for the BPS states are expressed by introducing parameters $\eta$ and $\bar{\eta}$ as 
\begin{align}
\zeta^{11}&=w_1\eta, & \zeta^{22}&=\bar{w}_{3}\bar{\eta}, & 
\zeta^{13}&=w_{2}\eta, & \zeta^{24}&=\bar{w}_{4}\bar{\eta},
\label{eq:1/4BPS_cond_VW}
\end{align}
and their complex conjugates.

Substituting \eqref{eq:1/4BPS_cond_VW} into the conditions $\delta \Lambda=0$ and $\delta \bar{\Lambda}=0$, we have the BPS equations. 
In order to obtain the BPS equations which have the nontrivial electric and magnetic fields, we must take the limit $w_1,w_3\to0$, $w_1,w_4\to0$, $w_2, w_3\to0$, or $w_2, w_4\to0$.
These four limits are shown to be equivalent to each other by the field redefinition.
Let us consider the limit $w_2, w_4\to0$.
More precisely we need to take the limit $w_2,w_4 \to 0$ such that 
\begin{gather}
\frac{w_2}{w_1} \to 0, \quad \frac{w_3}{w_1} \to - e^{i \theta}, \quad
 \frac{w_4}{w_1} \to 0,
\end{gather}
are satisfied.
Here the angle parameter $\theta$ is given by
\begin{align}
e^{i\theta}=\frac{i(q_1+iq_2)}{|q_1+iq_2|}. 
\label{eq:theta_1/4VW}
\end{align}
This can be done by the limit 
$q_5, q_6 \to 0$
 with fixed 
$q_5/q_6$, 
$q_1$ and $q_2$.
The preserved supercharges now have the same form as \eqref{eq:preserved_Q_half}.
The deformed 1/4 BPS dyon equations are 
\begin{align}
& E_i -(D_i\phi_2+\omega^m_2F_{m i})=0, \notag \\
& B_i +( D_i\phi_1+\omega^m_1F_{m i} )-i\delta_{3,i}\left([\varphi_3, \varphi_4] - [\varphi_5, \varphi_6]-i\Omega^m_5D_m\varphi_6+i\Omega^m_6D_m\varphi_5 \right)=0, \notag \\
& D_0\phi_1+\omega^m_1F_{m0} -i[\phi_1, \phi_2]+\omega^m_1D_m\phi_2-\omega^m_2D_m\phi_1=0, \notag \\
& D_0\phi_2 +\omega^m_2F_{m0} = 0, \notag \\
& (D_1+iD_2)(\varphi_3-i\varphi_4)=0,\notag \\
& D_3\varphi_a+i(-1)^{f(a)}\Bigl([\phi_1,\varphi_{f(a)}]+i\omega^m_1D_m\varphi_{f(a)}-i(-1)^{f(a)}\varepsilon^1_1\varphi_a \Bigr)=0,\notag \\
& D_0\varphi_a+i[\phi_2,\varphi_a]-\omega^m_2D_m\varphi_a+(-1)^{f(a)}\varepsilon^1_1\varphi_{f(a)}=0, \quad (a=3,4), \notag \\
& (D_1+iD_2)(\varphi_5+i\varphi_6)+(\Omega_5^m+i\Omega^m_6)(F_{m 1}+iF_{m 2})=0,\notag \\
& D_3\varphi_b+\Omega^m_bF_{m3}+i(-1)^{f(b)}\Bigl([\phi_1,\varphi_{f(b)}]+i\omega^m_1D_m\varphi_{f(b)}-i\Omega^m_{f(b)}D_m\phi_2 \Bigr)=0,\notag \\
& D_0\varphi_b+\Omega^m_bF_{m 0}+i[\phi_2, \varphi_b]-\omega^m_2D_m\varphi_b+\Omega^m_bD_m\phi_2=0, \quad (b=5,6), \notag \\
& [\varphi_3, \varphi_6]-[\varphi_4,\varphi_5]-i\Omega_6^m D_m\varphi_3+i\Omega^m_5D_m\varphi_4-i\epsilon^1_5\varphi_3-i\epsilon^1_6\varphi_4=0, \notag \\
& [\varphi_3, \varphi_5]+[\varphi_4, \varphi_6]-i\Omega^m_5D_m\varphi_3-i\Omega^m_6D_m\varphi_4+i\epsilon^1_6\varphi_3-i\epsilon^1_5\varphi_4=0,
\label{eq:quarterBPSeqVW}
\end{align}
where $\phi_a$, $\omega_a^m$ and $\varepsilon_a^1$ ($a=1,2$) are given by \eqref{para-rot}, and the functions $f(a)$ and $g(a)$ are given by \eqref{eq:def_fandg}.
The equations \eqref{eq:quarterBPSeqVW} are the $\Omega$-deformation of the undeformed 1/4 BPS equation \eqref{quarterdyonequndef}.
They become the same as in the half twist \eqref{eq:BPSeqhalf} when $\epsilon_5^1=\epsilon_6^1=0$ and $\mu_a=\varepsilon_a^1/2$ ($a=1,2$) \cite{Ito:2013eva}.
We have to impose the Gauss' law constraint in the NS limit:
\begin{align}
 & D_iE_i - \sum_{a=1,2,5,6}\biggl( D^m[\Omega_{ma}(D_0\varphi_a+\Omega^p_aF_{p0})] - i[\varphi_a, D_0\varphi_a+\Omega^p_aF_{p0}] \biggr) +i\sum_{a=3,4}[\varphi_a, D_0\varphi_a]=0,
\end{align}
for the consistency with the equations of motion.


\subsection{BPS equations from Bogomol'nyi completion}

We can derive the deformed BPS equations \eqref{eq:BPSeqhalf}, \eqref{eq:halfBPSeqVW} and \eqref{eq:quarterBPSeqVW} from the Bogomol'nyi completion of the
energy $E=P^{0}$, which is given by 
\begin{align}
E = \frac{1}{\kappa g^2} \int \! d^3 x \ 
\mathrm{Tr}
\left[
\frac{1}{2} E_i^2 + \frac{1}{2} B_i^2 
+ \frac{1}{2} G_{0a}^2 
+ \frac{1}{2} G_{ia}^2
+ \frac{1}{4} H_{ab}^2
\right].
\label{eq:energy}
\end{align}
Here the fermionic part is omitted. 
We will study the completion of the energy for each BPS equation.

\paragraph{1/2 BPS equations in the half twist}
We first perform the Bogomol'nyi completion of the energy in the
half twist and will derive the 1/2 BPS equations \eqref{eq:BPSeqhalf}.
We find that the energy \eqref{eq:energy} is 
rewritten as the sum of the complete squared forms, the magnetic charges
and the deformed electric charges:
\begin{align}
E =& \ \int \! d^3 x \ \frac{1}{\kappa g^2} 
\mathrm{Tr} 
\Big[
\frac{1}{2} 
\left\{
E_i - (D_i \phi_2 + \omega_2^j F_{ji})
\right\}^2 +
\frac{1}{2} 
\left\{
B_i + (D_i \phi_1 + \omega_1^j F_{ji}) -
 \delta_{3i} (H_{34} - H_{56} )
\right\}^2
\notag \\
& \ 
\qquad \qquad \qquad +
\frac{1}{2} 
\left(
D_0 \phi_1 + \omega_1^j F_{j0} - h_{12} 
\right)^2
+
\frac{1}{2} 
\left(
D_0 \phi_2 + \omega_2^j F_{j0}
\right)^2
\notag \\
& \ 
\qquad \qquad \qquad +
\frac{1}{2} | (D_1 + i D_2) (\varphi_3 - i \varphi_4) |^2
+
\frac{1}{2} | (D_1 - i D_2) (\varphi_5 - i \varphi_6) |^2
\notag \\
& \ 
\qquad \qquad \qquad 
+
\frac{1}{2} (D_3 \varphi_3 + h_{14} )^2
+
\frac{1}{2} (D_3 \varphi_4 - h_{13} )^2
+
\frac{1}{2} (D_3 \varphi_5 - h_{16} )^2
+
\frac{1}{2} (D_3 \varphi_6 + h_{15} )^2
\notag \\
& \ 
\qquad \qquad \qquad 
+
\frac{1}{2} \sum_{a=3,4,5,6} (D_0 \varphi_a + h_{2a} )^2
+
\frac{1}{2} (H_{36} - H_{45})^2
+ 
\frac{1}{2} (H_{35} + H_{46} )^2
\Big]
\notag \\
& \ - (q_1^{(e)} + q_2^{(m)}) \sin \theta + 
(q_2^{(e)} - q_1^{(m)}) \cos \theta.
\label{eq:hthbps}
\end{align}
Here we define
\begin{align}
 h_{1a} & \equiv H_{1a}\cos\theta + H_{2a}\sin\theta, & 
h_{2a} & \equiv -H_{1a}\sin\theta + H_{2a}\cos\theta.
\label{eq:def_hatandcheck}
\end{align}
Then the energy is bounded from below as 
\begin{align}
E \ge - (q_1^{(e)} + q_2^{(m)}) \sin \theta + 
(q_2^{(e)} - q_1^{(m)}) \cos \theta.
\end{align}
In order that this inequality holds for any $\theta$, 
the energy must satisfy the following inequality:
\begin{align}
E \ge \sqrt{
(q_1^{(e)} + q_2^{(m)})^2 + (q_2^{(e)} - q_1^{(m)})^2
}.
\label{eq:energy_bound_ht}
\end{align}
The inequality is saturated when each squared form in
\eqref{eq:hthbps} vanishes and $\theta$ is given by \eqref{eq:theta}.
When the equality holds, we obtain the 1/2 BPS equations \eqref{eq:BPSeqhalf}.
The energy bound \eqref{eq:energy_bound_ht} is given by the absolute
value of the central charge \eqref{eq:BPSbound_half}.

\paragraph{1/2 BPS equations in the Vafa-Witten and the Marcus twists}
We will derive the 1/2 BPS equations \eqref{eq:halfBPSeqVW} in the
Vafa-Witten and Marcus twists from the Bogomol'nyi completion of the energy.
We find that the energy \eqref{eq:energy} is rewritten as
\begin{align}
E =& \ \int \! d^3 x \ \frac{1}{\kappa g^2} 
\mathrm{Tr} \
\biggl[
\frac{1}{2} 
\left\{
E_i+(D_i\varphi_1+\Omega_1^jF_{ji})\sin\theta
\right\}
^2
\notag \\
& \ 
+
\frac{1}{2} 
\left\{
B_i+(D_i\varphi_1+\Omega_1^jF_{ji})\cos\theta-\delta_{3,i}H_{34}
\right\}
^2
+
\frac{1}{2}
(D_0\varphi_1+\Omega_1^jF_{j0})^2
\notag \\
& \
+
\frac{1}{2}
\sum_{\begin{subarray}{c}i=1,2 \\ a=2,5,6\end{subarray}} (D_i\varphi_a+\Omega_a^jF_{ji})^2
+
\frac{1}{2}
\sum_{\begin{subarray}{c}m=0,3 \\ a=2,5,6\end{subarray}} (D_m\varphi_a+\Omega_a^jF_{jm})^2\cos^2\theta
\notag \\
& \
+
\frac{1}{2}
\sum_{a=2,5,6} 
\left\{
(D_0\varphi_a+\Omega_a^jF_{j0})\sin\theta - H_{1a}
\right\}^2
\notag \\
& \ 
+
\frac{1}{2}
\sum_{a=2,5,6}
\left\{
(D_3\varphi_a+\Omega_a^jF_{j3})\sin\theta + \frac{1}{2}\varepsilon^{abc}H_{bc}
\right\}^2
+
\frac{1}{2}| (D_1 + i D_2) (\varphi_3 - i \varphi_4) |^2
\notag \\
& \ 
+
\frac{1}{2}
\sum_{a=3,4} 
\left\{
(D_3 \varphi_a + (-1)^{f(a)} H_{1,f(a)} \cos\theta)^2
+
(D_0 \varphi_a - H_{1a} \sin\theta)^2
\right\}
\notag \\
& \ 
+ 
\frac{1}{2} (H_{23}^2+H_{24}^2+H_{35}^2+H_{36}^2+H_{45}^2+H_{46}^2)
\biggr]
- q_1^{(e)} \sin \theta - 
 q_1^{(m)} \cos \theta,
\label{eq:vwthbps}
\end{align}
where 
$\theta$ is a phase factor and 
$\varepsilon^{abc}$ ($a,b,c=2,5,6$) is the totally antisymmetric tensor
with $\varepsilon^{256}=1$.
$f(a)$ is given by \eqref{eq:def_fandg}.
Maximizing the energy bound with respect to $\theta$, we obtain the
inequality
\begin{align}
E \ge  \sqrt{ (q_{1}^{(m)})^2 + (q_{1}^{(e)})^2}.
\label{eq:energy_bound_VW}
\end{align}
The inequality is saturated when each squared form in
\eqref{eq:vwthbps} vanishes and $\theta$ satisfies $\tan \theta =
q_1^{(e)}/q_1^{(m)}$, which is the same as \eqref{eq:theta1/2BPSVW}.
The condition that each squared form vanishes gives the 1/2 BPS equations \eqref{eq:halfBPSeqVW}.
The energy bound \eqref{eq:energy_bound_VW} is given by 
the BPS bound of \eqref{eq:BPSbound_VW} 
for $q_{a}^{(m)}=q_{a}^{(e)}=0$ $(a=2,5,6)$, which corresponds to the condition such that the supercharges \eqref{preserved_1/2BPS} are preserved.

\paragraph{1/4 BPS equations in the Vafa-Witten and the Marcus twists}
We next discuss the 1/4 BPS equations \eqref{eq:quarterBPSeqVW}
from the Bogomol'nyi completion.
We find that the energy \eqref{eq:energy} is rewritten as 
\begin{align}
E =& \ \int \! d^3 x \ \frac{1}{\kappa g^2} \mathrm{Tr}
\Big[
\frac{1}{2} 
\left\{
E_i - (D_i \phi_2 + \omega_2^j F_{ji})
\right\}^2
+
\frac{1}{2} 
\left\{
B_i + (D_i \phi_1 + \omega_1^j F_{ji}) -
 \delta_{3i} (H_{34} - H_{56})
\right\}^2
\notag \\
& \
\qquad \qquad \qquad 
+
\frac{1}{2} 
\left\{
D_0 \phi_1 + \omega_1^j F_{j0} - h_{12}
\right\}^2
+
\frac{1}{2} (D_0 \phi_2 + \omega_2^j F_{j0})^2
\notag \\
& \ 
\qquad \qquad \qquad 
+
\frac{1}{2} | (D_1 + i D_2) (\varphi_3 - i \varphi_4) |^2
\notag \\
& \ 
\qquad \qquad \qquad 
+
\frac{1}{2} (D_3 \varphi_3 + h_{14})^2
+
\frac{1}{2} (D_3 \varphi_4 - h_{13})^2
\notag \\
& \ 
\qquad \qquad \qquad 
+
\frac{1}{2} | (D_1 - i D_2) (\varphi_5 - i \varphi_6) 
+ (F_{j1} - i F_{j2}) (\Omega_5^j - i \Omega_6^j ) |^2
\notag \\
& \ 
\qquad \qquad \qquad 
+
\frac{1}{2} (D_3 \varphi_5 + \Omega_5^j F_{j3} - h_{16} )^2
+
\frac{1}{2} (D_3 \varphi_6 + \Omega_6^j F_{j3} + h_{15} )^2
\notag \\
& \ 
\qquad \qquad \qquad 
+
\frac{1}{2} \sum_{a=3,4} (D_0 \varphi_a + h_{2a})^2
+
\frac{1}{2} \sum_{a=5,6} (D_0 \varphi_a + \Omega^j_a F_{j0} + h_{2a})^2
\notag \\
& \ 
\qquad \qquad \qquad 
+
\frac{1}{2} (H_{36} - H_{45})^2 + \frac{1}{2} (H_{35} + H_{46})^2
\Big]
\notag \\
& \ - (q_1^{(e)} + q_2^{(m)}) \sin \theta + (q_2^{(e)} - q_1^{(m)}) \cos
 \theta.
\label{eq:vwtqbps}
\end{align}
Then the energy is bounded from below as
\begin{align}
E \ge \sqrt{
(q_1^{(e)} + q_2^{(m)})^2 + (q_2^{(e)} - q_1^{(m)})^2
}.
\label{eq:energy_bound_vw2}
\end{align}
The inequality is saturated when $\theta$ is given by 
\eqref{eq:theta_1/4VW} and the complete squared forms in
\eqref{eq:vwtqbps} vanish. 
The latter condition gives the 1/4 BPS equations \eqref{eq:quarterBPSeqVW}.
The energy bound \eqref{eq:energy_bound_vw2} is given by the 
BPS bound \eqref{eq:BPSbound_VW}
with $q_{a}^{(m)}=q_{a}^{(e)}=0$ $(a=5,6)$. 


\section{Conclusion}
In this paper, we studied 
the central charge of the supersymmetry algebra and the deformed BPS dyon equations
in the NS limit of four-dimensional $\mathcal{N} = 4$ super Yang-Mills theory in the
$\Omega$-background.
We took the NS limit such that the Poincar\'e symmetry in 
the $(x^0,x^3)$-subspace is recovered and the energy and the momentum along the $x^3$-direction of the theory are conserved.

The supersymmetry transformation of the supercurrent in the general
ten-dimensional curved space provides 
the central charges of the supersymmetry algebra.
By the dimensional reduction to four dimensions, we 
obtained the central charge of the $\mathcal{N} = 4$ super Yang-Mills
theory in the $\Omega$-background
associated with the topological twists of ${\cal N}=4$ supersymmetry.
The central charge  is given by the electric and the magnetic
charges. As in the ${\cal N}=2$ case, the magnetic charge formula is not deformed by the $\Omega$-background, 
while the electric charge formula is deformed.
We obtained the BPS bound for the mass from the supersymmetry algebras.
By the condition that the preserved supersymmetry transformation of the fermions
vanish, we obtained the BPS equations for dyons.
We found 1/2 BPS dyon equations for the half, the Vafa-Witten and the Marcus twists and 1/4 BPS dyon equations in the Vafa-Witten and the Marcus twists.
These equations have been also derived from the Bogomol'nyi completion of the $\Omega$-deformed energy.

It is an interesting problem to study the solutions of the deformed BPS
equations and the deformed BPS spectrum.
Since the $\Omega$-background in the NS limit breaks the Poincar\'e symmetry in $(x^1,x^2)$-subspace, 
the BPS equations for dyons do not allow three-dimensional spherically symmetric solutions. 
It rather admits an axially symmetric solutions.
For the $\mathcal{N} = 2$ super Yang-Mills theory in the $\Omega$-background, the BPS monopole equations and their axially symmetric solutions with unit charge have been found \cite{Ito:2011ta}.
Since the present BPS equations reduce to the ${\cal N}=2$
ones by the projection, we expect that for the $\mathcal{N}=4$ theory the equations have similar
type of solutions.
It would be interesting to study the Nahm construction of the monopoles \cite{Nahm:1979yw} for the construction of the solutions with higher charges and their moduli space.
It is also interesting to study S-duality 
\cite{Hellerman:2012rd, Lambert:2013lxa, Lambert:2014fma,
Billo':2015ria}
of the $\Omega$-deformed ${\cal N}=4$ theory as well as the
relation to the integrable systems \cite{Nekrasov:2009rc,Nekrasov:2010ka,Bulycheva:2012ct}.

\subsection*{Acknowledgements} 
The work of K.~I.  is supported 
in part by Grant-in-Aid for Scientific Research from the Japan Ministry of Education, 
Culture, Sports, Science and Technology. 
The work of H.~N. is supported in part by the NSFC 
(Grant No. 11175039 and 11375121).
The work of S.~S is supported in part by Kitasato University Research
Grant for Young Researchers.

\begin{appendix}

\section{Four- and six-dimensional Dirac matrices} \label{sc:appA}
In this appendix, we provide the decomposition of the ten-dimensional
Dirac matrices by the dimensional reduction to four dimensions and introduce 
the four and the six-dimensional Dirac matrices.
The local Lorentz group $SO(10)$ is reduced to $SO(4) \times SO(6)_I$, 
where $SO(4) \simeq SU(2)_L\times SU(2)_R$ is the Lorentz group
in four dimensions and $SO(6)_I \simeq SU(4)_I$ becomes the R-symmetry group.
We decomposed the ten-dimensional vector index $M$ as  
$M=(m,a+4)$, $(m=1,\ldots,4,\ a=1,\ldots,6)$.
Here
$m,a$ are the indices for four-dimensional spacetime and the
six-dimensional internal space. 
The ten-dimensional Dirac matrices are decomposed as 
\begin{align}
\Gamma^m = -i 
\begin{pmatrix} 
0 & (\sigma^{m})_{\alpha\dot{\alpha}} \\ 
 (\bar{\sigma}^{m})^{\dot{\alpha}\alpha} & 0 
\end{pmatrix}
\otimes \boldsymbol{1}_{8}
, \qquad 
\Gamma^{a+4} = 
\begin{pmatrix} 
\boldsymbol{1}_{2} & 0 \\ 0 & -\boldsymbol{1}_{2} 
\end{pmatrix}
\otimes 
\begin{pmatrix} 
0 & 
(\Sigma_{a})^{AB}
 \\ (
\bar{\Sigma}_{a})_{AB}
 & 0 
\end{pmatrix},
\end{align}
where 
$(\sigma^m)_{\alpha \dot{\alpha}}$,
$(\bar{\sigma}^m)^{\dot{\alpha} \alpha}$, $(\Sigma_a)^{AB} \ (\bar{\Sigma}_a)_{AB}$ are
the four- and the six-dimensional Dirac matrices defined below.
The indices $\alpha, \dot{\alpha}=1, 2$ are the $SU(2)_L$ and $SU(2)_R$
 spinor indices, respectively. These indices are raised and lowered by the antisymmetric
 $\varepsilon$-symbol normalized as $\varepsilon^{12}=-\varepsilon_{12}=1$.
$A=1,2,3,4$ is the index for the fundamental representation of $SU(4)_I$.

The four-dimensional Dirac matrices $(\sigma^m)_{\alpha\dot{\alpha}}$
and $(\bar{\sigma}^{m})^{\alpha\dot{\alpha}}$ with the Euclidean signature are
$\sigma^{m} = (i \tau^1, i \tau^2, i \tau^3, \mathbf{1}_2)$,
$\bar{\sigma}^{m} = (- i \tau^1, - i \tau^2, - i \tau^3, \mathbf{1}_2)$, $(m,n = 1,2,3,4)$.
The four-dimensional Dirac matrices in the Minkowski signature are
$\sigma^{m} = (-\mathbf{1}_2,  \tau^1,  \tau^2,  \tau^3)$,
$\bar{\sigma}^{m} = (-\mathbf{1}_2, - \tau^1, - \tau^2, - \tau^3)$,
$(m,n = 0,1,2,3)$. 
Here $\boldsymbol{1}_{n}$ denotes the 
$n\times n$ identity matrix and $\tau^i \ (i = 1,2,3)$ are the
Pauli matrices. 
The four-dimensional Lorentz generators 
are defined by 
$\sigma^{mn} = \frac{1}{4} (\sigma^{m} \bar{\sigma}^{n} -
\sigma^{n} \bar{\sigma}^{m})$, $\bar{\sigma}^{mn} = \frac{1}{4}
(\bar{\sigma}^{m} \sigma^{n} - \bar{\sigma}^{n} \sigma^{m})$.

The six-dimensional Dirac matrices 
$(\Sigma_{a})^{AB}$ and $(\bar{\Sigma}_{a})_{AB}$ are defined by 
\begin{align}
\Sigma_{1}
&=
\begin{pmatrix} i\tau^{2} & 0 \\ 0 & i\tau^{2} \end{pmatrix},
&
\Sigma_{2}
&=
\begin{pmatrix} \tau^{2} & 0 \\ 0 & -\tau^{2} \end{pmatrix},
&
\Sigma_{3}
&=
\begin{pmatrix} 0 & -\tau^{3} \\ \tau^{3} & 0 \end{pmatrix},
\notag\\[2mm]
\Sigma_{4}
&=
\begin{pmatrix} 0 & i\boldsymbol{1}_{2} \\ -i\boldsymbol{1}_{2} & 0 
\end{pmatrix},
&
\Sigma_{5}
&=
\begin{pmatrix} 0 & - \tau^{1} \\ \tau^{1} & 0 \end{pmatrix},
&
\Sigma_{6}
&=
\begin{pmatrix} 0 & \tau^{2} \\ \tau^{2} & 0 \end{pmatrix},
\notag\\[2mm]
\bar{\Sigma}_{1}
&=
\begin{pmatrix} -i\tau^{2} & 0 \\ 0 & -i\tau^{2} \end{pmatrix},
&
\bar{\Sigma}_{2}
&=
\begin{pmatrix} \tau^{2} & 0 \\ 0 & -\tau^{2} \end{pmatrix},
&
\bar{\Sigma}_{3}
&=
\begin{pmatrix} 0 & \tau^{3} \\ -\tau^{3} & 0 \end{pmatrix},
\notag\\[2mm]
\bar{\Sigma}_{4}
&=
\begin{pmatrix} 0 & i\boldsymbol{1}_{2} \\ -i\boldsymbol{1}_{2} & 0 
\end{pmatrix},
&
\bar{\Sigma}_{5}
&=
\begin{pmatrix} 0 &  \tau^{1} \\ - \tau^{1} & 0 \end{pmatrix},
&
\bar{\Sigma}_{6}
&=
\begin{pmatrix} 0 & \tau^{2} \\ \tau^{2} & 0 \end{pmatrix}.
\end{align}
The Lorentz generators 
$(\Sigma_{ab})^A{}_B$
 and
 $(\bar{\Sigma}_{ab})_A{}^B$
 are defined by
\begin{align}
 \Sigma_{ab}=\frac{1}{4}( \Sigma_a \bar{\Sigma}_b - \Sigma_b \bar{\Sigma}_a), 
\quad \bar{\Sigma}_{ab}=\frac{1}{4}(\bar{\Sigma}_a \Sigma_b - \bar{\Sigma}_b \Sigma_a).
\end{align}
We also define 
the antisymmetrized products of the six-dimensional Dirac matrices $(\Sigma_{abc})^{AB}$, $(\bar{\Sigma}_{abc})_{AB}$, $(\Sigma_{abcd})^A{}_B$ and $(\bar{\Sigma}_{abcd})_A{}^B$ by
\begin{align}
\Sigma_{abc} & = \Sigma_{[a} \bar{\Sigma}_b \Sigma_{c]}, \qquad 
 \bar{\Sigma}_{abc} = \bar{\Sigma}_{[a} \Sigma_b \bar{\Sigma}_{c]},
\notag \\
\Sigma_{abcd} & = \Sigma_{[a} \bar{\Sigma}_b \Sigma_c \bar{\Sigma}_{d]},
 \qquad 
\bar{\Sigma}_{abcd} = \bar{\Sigma}_{[a} \Sigma_{b} \bar{\Sigma}_{c} \Sigma_{d]},
\end{align}
where the square bracket denotes the antisymmetrization defined in subsection \ref{sc:half}.

\section{Vacuum conditions} \label{sc:appB}
In this appendix, we study the vacuum structure of the $\mathcal{N} =
4$ super Yang-Mills theory in the $\Omega$-background.
This is necessary to determine the surface terms in the integral of the 
central charges.
The vacua of the theory are defined by the conditions $F_{mn} =
G_{ma} = H_{ab} =0$.
From the conditions $F_{mn} =0$ and $D_{m} \varphi_a = 0$, 
the gauge field $A_{m}$ vanishes and the scalar fields $\varphi_a$
take the constant values $\langle \varphi_a \rangle$ up to gauge transformation.
The values of the scalar fields are determined by the condition
$H_{ab} = 0$ which reads
\begin{align}
i [\langle \varphi_a \rangle, \langle \varphi_b \rangle] - T_{ab} {}^c
 \langle \varphi_c \rangle =0.
\label{eq:vacuumcond}
\end{align}
We solve the vacuum condition \eqref{eq:vacuumcond} 
for the $\Omega$-backgrounds associated with the
topological twists.

\paragraph{Half twist}
In the case of the half twist, the non-zero components of $T_{ab} {}^c$
are
\begin{align}
& T_{13} {}^4 = - T_{14} {}^3 = \frac{i}{4} (\epsilon^1_1 + \epsilon^2_1
 + 2 m_1) , \qquad 
T_{15} {}^6 = - T_{16} {}^5 = - \frac{i}{4} (\epsilon^1_1 + \epsilon^2_1 - 2 m_1),
\notag \\
& T_{23} {}^4 = - T_{24} {}^3 = \frac{i}{4} (\epsilon^1_2 + \epsilon^2_2
 + 2 m_2), \qquad 
T_{25} {}^6 = - T_{26} {}^5 = - \frac{i}{4} (\epsilon^1_2 + \epsilon^2_2 - 2 m_2).
\label{eq:G_half}
\end{align}
From the vacuum condition \eqref{eq:vacuumcond} with the torsion 
\eqref{eq:G_half}, we find that 
$\langle \varphi_a \rangle \ (a=1,2)$ belong to the Cartan subalgebra of the
Lie algebra of the gauge group and $\langle \varphi_a \rangle = 0 \
(a=3,\ldots,6)$.
The vacuum solution does not change in the NS limit.

\paragraph{Vafa-Witten twist}
In the case of the Vafa-Witten twist, the non-zero components of $T_{ab} {}^c$ 
are
\begin{align}
& T_{13} {}^4 = - T_{14} {}^3 = \frac{i}{2} (\epsilon^1_1 + \epsilon^2_1), \qquad
  T_{23} {}^4 = - T_{24} {}^3 = \frac{i}{2} (\epsilon^1_2 +
 \epsilon^2_2), \notag \\
& T_{53} {}^4 = - T_{54} {}^3 =  \frac{i}{2} (\epsilon^1_5 + \epsilon^2_5), \qquad
  T_{63} {}^4 = - T_{64} {}^3 =  \frac{i}{2} (\epsilon^1_6 + \epsilon^2_6).
\label{eq:G_VW}
\end{align}
From the vacuum condition \eqref{eq:vacuumcond} with the torsion \eqref{eq:G_VW},
$\langle \varphi_a \rangle \ (a=1,2,5,6)$ belong to the Cartan subalgebra and $\langle \varphi_a \rangle = 0 \ (a=3,4)$.
The vacuum solution does not change in the NS limit.

\paragraph{Marcus twist}
In the case of the Marcus twist, the non-zero components of $T_{ab} {}^c$ are
\begin{align}
& T_{13} {}^4 = - T_{14} {}^3 = \frac{i}{2} \epsilon^1_1, \quad 
  T_{15} {}^6 = - T_{16} {}^5 = - \frac{i}{2} \epsilon^2_1, \notag \\
& T_{23} {}^4 = - T_{24} {}^3 = \frac{i}{2} \epsilon^1_2, \quad
  T_{25} {}^6 = - T_{26} {}^5 = - \frac{i}{2} \epsilon^2_2.
\label{eq:G_Marcus}
\end{align}
From the vacuum condition \eqref{eq:vacuumcond} and the torsion \eqref{eq:G_Marcus},
$\langle \varphi_a \rangle \ (a=1,2)$ belong to the Cartan subalgebra
and $\langle \varphi_a \rangle = 0 \ (a=3, \ldots,6)$.
We note that in the NS limit $\epsilon^2_a \to 0$, 
$\langle\varphi_{5}\rangle$ and $\langle\varphi_{6}\rangle$ also belong to
the Cartan subalgebra.
This NS limit corresponds to the special case of the Vafa-Witten twist.
\end{appendix}

\end{document}